%% file: main_FINAL.tex
\documentclass[final,1p,times]{elsarticle}

\usepackage[autonum,colorhypersetup]{shortex}

\usepackage{graphicx}
\graphicspath{{figures/}}

\usepackage{acro}
\acsetup{single={1},	
	uppercase/list}
	
\input{./acronyms} 

\usepackage{lineno}

\journal{arXiv}

\begin{document}

\begin{frontmatter}

\title{A view on learning robust goal-conditioned value functions: Interplay between RL and MPC\tnoteref{label1}}

\tnotetext[label1]{Please cite the journal version in Annual Reviews in Control \url{https://doi.org/10.1016/j.arcontrol.2025.101027}}

\author[ucb]{Nathan P. Lawrence}
\ead{nplawrence@berkeley.edu}
\author[ubc_math]{Philip D. Loewen}
\author[honeywell]{Michael G. Forbes}
\author[ubc_chbe]{R. Bhushan Gopaluni}
\author[ucb]{Ali Mesbah}
\ead{mesbah@berkeley.edu}

\address[ucb]{Department of Chemical and Biomolecular Engineering, University of California, Berkeley, CA 94720, USA}
\address[ubc_math]{Department of Mathematics, University of British Columbia, Vancouver, BC V6T 1Z2, Canada}
\address[ubc_chbe]{Department of Chemical and Biological Engineering, University of British Columbia,
Vancouver, BC V6T 1Z3, Canada}
\address[honeywell]{Honeywell Process Solutions, North Vancouver, BC V7J 3S4, Canada}


\begin{abstract}
\Ac{RL} and \ac{MPC} offer a wealth of distinct approaches for automatic decision-making under uncertainty.
Given the impact both fields have had independently across numerous domains, there is growing interest in combining the general-purpose learning capability of \ac{RL} with the safety and robustness features of \ac{MPC}.
To this end, this paper presents a tutorial-style treatment of \ac{RL} and \ac{MPC}, treating them as alternative approaches to solving \aclp{MDP}.
In our formulation, \ac{RL} aims to learn a \emph{global} value function through offline exploration in an uncertain environment, whereas \ac{MPC} constructs a \emph{local} value function through online optimization.
This local-global perspective suggests new ways to design policies that combine robustness and goal-conditioned learning.
Robustness is incorporated into the \ac{RL} and \ac{MPC} pipelines through a scenario-based approach.
Goal-conditioned learning aims to alleviate the burden of engineering a reward function for \ac{RL}.
Combining the two leads to a single policy that unites a robust, high-level \ac{RL} terminal value function with short-term, scenario-based \ac{MPC} planning for reliable constraint satisfaction.
This approach leverages the benefits of both \ac{RL} and \ac{MPC}, the effectiveness of which is demonstrated on classical control benchmarks.
\end{abstract}
\end{frontmatter}

\section{Introduction}
\acresetall

\Ac{RL} and \ac{MPC} are optimization-based frameworks for decision-making \citep{bertsekas2022lessons}.
Model-free \ac{RL} represents a \emph{sample-based} approach in which a control policy is improved through trial and error in an uncertain environment \citep{sutton2018ReinforcementLearning}.
On the other hand, \ac{MPC} is a \emph{systems-based} approach in which forecasts are used to select appropriate control actions \citep{borrelli2017predictive}.
Both can be understood in the context of \acp{MDP}, but have enjoyed practical success in vastly different domains \citep{forbes2015ModelPredictive,mesbah2018Stochasticmodel,busoniu2018ReinforcementLearning,lawrence2024MachineLearning}.

\begin{figure}
	\includegraphics[width=3.3in]{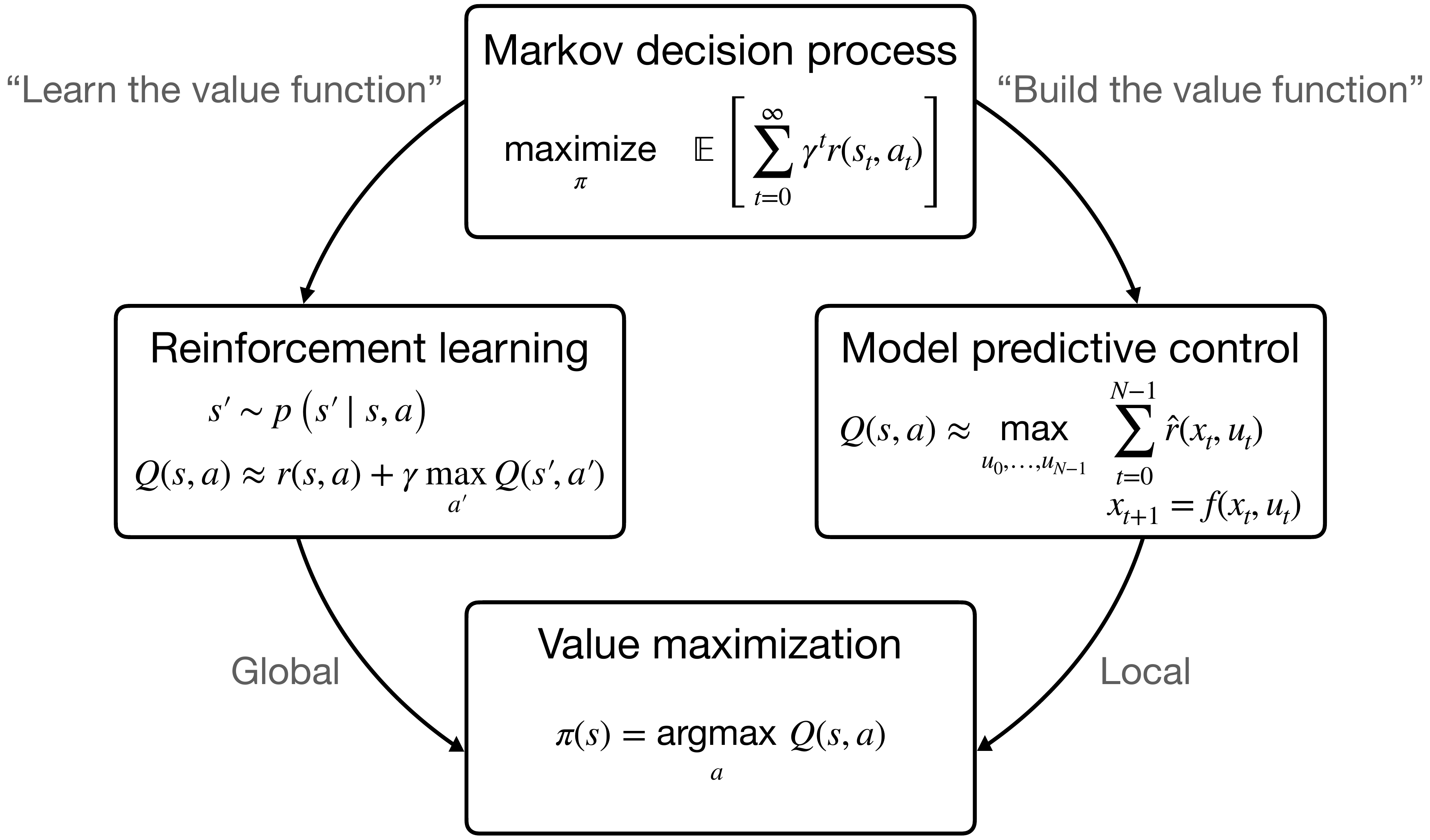}
	\caption{\Ac{RL} and \ac{MPC} can be seen as alternative approaches to solving \acp{MDP}. However, they both leverage the idea of selecting actions by maximizing a value function $Q$. The \ac{RL} agent learns a global value function offline, while the \ac{MPC} constructs a local value function for online control. $\hat{r}$ represents a tractable reward for the \ac{MPC} agent, possibly different from the true reward signal $r$.}
	\label{fig:mdp}
\end{figure}

Within the setting of \acp{MDP}, \ac{RL} and \ac{MPC} can be connected through the idea of value functions \citep{bertsekas2022lessons, bertsekas1996neuro}, a mechanism for predicting future performance; see \cref{fig:mdp}\footnote{Readers familiar with \ac{MPC} can recover the usual minimization problem by thinking of $-\hat{r}$ as the stage cost. We cast \ac{MPC} as a maximization problem for consistency within the overall framework and because $\hat{r}$ can be more general than a traditional stage cost.} for a conceptual diagram and \cref{sec:related} for an overview of prior work.
However, two challenges emerge in acquiring such a value function:
\begin{enumerate}
	\item {\bf Unknown cost.}\quad Desirable performance is often difficult to quantify precisely. \Ac{MPC} typically uses quadratic cost functions because they are tractable and a stability theory is available, but the parameters in the objective are only indirectly linked with appropriate closed-loop behavior. Fine-tuning is often required. \Ac{RL}, on the other hand, can learn from reward signals that express operational goals quite succinctly, such as a ``yes/no'' stimulus, but may require many trials to capture the designer's intent.  
	\item {\bf Unknown dynamics.}\quad A hallmark of \ac{RL} is its model-free learning capability, enabling it to generate a high-performing policy without a model of the system being controlled. \ac{MPC}, of course, requires a reasonably accurate system model. Since real-world environments are never truly stationary or precisely known, robust approaches to learning and modeling are essential.
\end{enumerate}

This paper addresses these two challenges by taking a fresh look at both \ac{RL} and \ac{MPC} from the vantage point of value function estimation.\footnote{Throughout, we refer to \emph{model-free \ac{RL}} and \emph{robust \ac{MPC}}, but simply state \ac{RL} and \ac{MPC} for brevity and generality.}
\Cref{sec:globalRL,sec:localMPC} present a classical overview of \ac{RL} and \ac{MPC} ideas:
\Iac{RL} agent explores its environment to synthesize a high-level value function from a reward signal.
This is a \emph{global} approach wherein, at deployment, the \ac{RL} agent simply queries its value-maximizing policy. 
On the other hand, \ac{MPC} represents a modular strategy in which prior physical knowledge and safety specifications are directly embedded in the form of equality and inequality constraints.
This results in a \emph{local} value structure characterized by the agent continually replanning online.

Taken together, we present a unified framework wherein the \ac{MPC} architecture can take advantage of an \ac{RL}-learned value function to calibrate its long-term cost predictions to the system of interest.
Conversely, the \ac{RL} agent benefits from the exact, constrained optimization of the \ac{MPC} module to produce safe actions.
Specialized techniques from both the \ac{RL} and \ac{MPC} literature are embraced to make training more efficient and action selection more robust.
In particular, this paper builds on the classical local-global view \citep{banker2025LocalGlobalLearning} by bringing together scenario-based planning and goal-conditioned learning into a single agent.
The contributions of this paper are summarized as follows:
\begin{enumerate}
	\item We use methods from robust \ac{MPC} to design actions guided by constraints and \iac{RL} terminal value function; and a scenario-based setup inspired by \ac{MPC} to train the robust \ac{RL} agent from simulated experience data.
	\item We use goal-conditioned \ac{RL} techniques to learn a high-level value function to augment the \ac{MPC} agent.
	\item We give a tutorial-style treatment of the local-global value function perspectives of \ac{MPC} and \ac{RL}. Moreover, we elaborate on the implementation of robust \ac{MPC} methods in \iac{RL} ecosystem.
\end{enumerate}
This paper serves as a value function-centric overview for researchers in either area of \ac{MPC} or \ac{RL}.
Value functions are fundamental to both areas, making them the ideal kernel for discussion.
Through this lens, we provide a balanced overview of key methodological advancements from both sides, towards interfacing \ac{RL} and \ac{MPC}.

\section{Learning a global value function through RL}
\label{sec:globalRL}

This section introduces \acp{MDP} and the \ac{RL} perspective for solving them.
Our key point is that \ac{RL} aims to produce a \emph{global} value function over the state-action space.
This contrasts with \ac{MPC}, detailed in \cref{sec:localMPC}, which builds a \emph{local} value function through the combination of constraints, costs, and replanning.

\subsection{Markov decision processes}
\label{sec:mdp}

We consider an optimization problem of the form:
\begin{equation}
\begin{aligned}
    &\underset{\pi}{\text{maximize}} && \mathbb{E}_{s_0, g \sim p\left(s_0, g\right)}\left[ V^{\pi}_g (s_0) \right].
\end{aligned}
\label{eq:abstractObj}
\end{equation}
The idea behind \cref{eq:abstractObj} is the following:
Starting from some \emph{state} $s_0$ in a dynamic environment, design a \emph{policy} $\pi$ that brings future states $s_1, s_{2}, \ldots$ to a desired \emph{goal} $g$.
The function $V^{\pi}_g$ indicates the \emph{value} of the policy; naturally, the ``best'' policy should act as efficiently as possible.

The construction and implementation of a policy are carried out by an \emph{agent}; the environment can be viewed as everything outside of the agent.
The environment involves a state space $\state$, while the agent selects actions in action space $\action$ according to the policy.
In the goal-conditioned setting, we also consider a goal space $\mathcal{G}$.\footnote{The introduction of a goal space generalizes the standard presentation of \acp{MDP} in which an agent has a single goal. While we interpret a goal as a desired state, goal representations may be more general, for example, embeddings representing winning positions in a game, or a target level of cumulative reward \citep{liu2022GoalConditionedReinforcement}. Additionally, this level of generality can be leveraged to enable planning over subgoals, often improving training speed \cite{chane-sane2021GoalConditionedReinforcement, lo2024GoalSpacePlanning}.}
For a fixed goal $g \in \mathcal{G}$ and a given state $s \in \state$, an action $a \in \action$ is applied to the environment, which produces a new state $s' \in \state$.
Successive states should eventually arrive at $g$.
We often index the states and actions in discrete time steps.
Starting from an initial state $s_0 \in \state$, we obtain a \emph{trajectory}
\[
\{ s_0, a_0, s_1, a_1, \ldots, s_t, a_t, s_{t+1}, \ldots \}.
\label{eq:traj}
\]
Crucially, we assume the state-action tuple $(s_t,a_t)$ completely characterizes the probability distribution over the next state $s_{t+1}$.
Informally, we are assuming that our predictions of $s_{t+1}$ based only on $(s_t,a_t)$ cannot be improved by including more of the history of the trajectory up to index $t$:
\[
\pp{p}{s_{t+1}}{s_t,a_t} = \pp{p}{s_{t+1}}{s_0, a_0, \ldots, s_t, a_t}.
\label{eq:markov}
\]
This is the so-called \emph{Markov property} wherein the transition dynamics distribution is characterized by conditional density $p$.

The environment dynamics encompass a large set of possible trajectories of the form shown in \cref{eq:traj}.
The desirability of each transition along a trajectory is summarized by a scalar-valued function $r_g: \state \times \action \to \reals$, known as the \emph{reward}.
Writing $r_t = r_g(s_t, a_t)$ produces a reward-annotated trajectory:
\[
\{ s_0, a_0, r_0, s_1, a_1, r_1 \ldots, s_t, a_t, r_t, s_{t+1}, \ldots \mid g \}.
\label{eq:sarsa}
\]
The trajectory is conditioned on $g$ to emphasize the goal-directed reward function.
A scalar value for a given trajectory can be assigned by specifying a constant $\gamma \in [ 0, 1 ]$ called the \emph{discount factor} and forming the discounted sum of future rewards:
\[
\sum_{t=0}^\infty \gamma^t r_g(s_t,a_t).
\label{eq:returnt0}
\]
If $\gamma = 0$, we interpret the series as $r_g(s_0,a_0)$; 
choosing $0<\gamma<1$ guarantees that the series converges (assuming the reward function is bounded) and assigns more weight to immediate rewards than to future rewards.

The link from states to actions is captured
in the \emph{policy}, a probability distribution over the set of actions that depends on the current state and the selected goal. For each state-goal pair $(s,g)$ in $\state \times \mathcal{G}$, $\pp{\pi}{a}{s,g}$ defines a conditional density at action $a \in \action$.
Each policy induces a probability on the set of trajectories mentioned above: Operationally, we focus on trajectories where the sample value of $s_{t+1}$ is determined by the density $\pp{p}{s_{t+1}}{s_t,a_t}$ after $a_t$ is drawn from the density $\pp{\pi}{a_t}{s_t,g}$.
Every policy $\pi$ assigns a scalar value to each point in the state space as follows:
\begin{equation}
	V^{\pi}_g \left( s \right) = \mathbb{E}_{\pi}\left[ \sum_{t=0}^{\infty} \gamma^{t} r_g (s_t,a_t) \middle| s_0 = s \right].
\label{eq:statevalue}
\end{equation} 
Here $V^\pi_g$ is a \emph{value function}: It returns the expected long-term reward accumulated under policy $\pi$ as a function of the trajectory's starting point.

In the context detailed above, the problem of determining a policy $\pi$ that maximizes the expected return over all possible trajectories from all possible starting states and goals $s_0, g \sim p\left(s_0, g\right)$ is a \emph{\acf{MDP}}.
In terse mathematical notation, our MDP is:
\begin{equation}
\begin{aligned}
    &\text{maximize} && J(\pi) = \mathbb{E}_{\pi}\left[ \sum_{t=0}^{\infty} \gamma^{t}r_g (s_t,a_t) \right]\\
    &\text{where} && a_t \sim \pp{\pi}{a_t}{s_t,g},\ s_{t+1} \sim \pp{p}{s_{t+1}}{s_t,a_t}\\
    & && s_0, g \sim p\left(s_0, g\right).
\end{aligned}
\label{eq:mdpobjective}
\end{equation}
Problem~(\ref{eq:mdpobjective}) is easy to state, but hard to solve.
Indeed, we cannot even evaluate the objective directly, as the infinite sum already restricts us to special cases or approximations.
Moreover, the transition probability distribution, which governs the system's dynamics, is generally treated as unknown.
Thus, the expectation is unavailable in closed form and must be estimated, for example, with empirical observations of the form in \cref{eq:sarsa}.
Finally, the space of all competing policies is intractable, meaning that some simply parameterized subset, such as that provided by a neural network, will have to suffice.
In what follows, we outline the \ac{RL} perspective for approximating $V^{\pi}_g$.

\subsection{The reinforcement learning approach}
\label{sec:RL}

\Ac{RL} offers an iterative, data-driven, and flexible framework for solving dynamic tasks:
\begin{itemize}
	\item \textbf{Iterative.}\quad Exact, analytical solutions are scarce. However, general optimality conditions, based on dynamic programming, inform elegant, iterative update schemes that improve decision-making performance over time. 
	\item \textbf{Data-driven.}\quad A model of the environment is not required (although one is welcome, if available). Instead, sequential data can be used in place of a dynamic model.
	\item \textbf{Flexible.}\quad The two aspects above mean \ac{RL} can be applied in many domains. Moreover, the training process is governed by a reward signal, which is a simple and intuitive way to impose goal-directed behavior.
\end{itemize}
This paper does not dwell on the minute details of individual algorithms, but rather looks to convey some general principles and structures that guide practical \ac{RL} solution methods.

\subsubsection{Evaluate, improve, and repeat...}

It is useful to define the state-action value function:
\begin{equation}
	Q^{\pi}_g \left( s, a \right) = \mathbb{E}_{\pi}\left[ \sum_{t=0}^{\infty} \gamma^{t} r_g (s_t, a_t) \middle| s_0 = s, a_0 = a \right].
\label{eq:stateactionvalue}
\end{equation} 
Given $Q^{\pi}_g$, one can obtain the state value as $V^{\pi}_g (s) = \mathbb{E}_{a \sim \pp{\pi}{a}{s,g}} \left[ Q^{\pi}_g (s, a) \right]$.\footnote{We often drop super/subscripts (or both) when we do not need to reference a specific policy or goal.}
Therefore, focusing on $Q$ is sufficient in light of our objective in \cref{eq:abstractObj}. This is beneficial due to the additional degree of freedom in the action component.
Indeed, if we had some oracle mapping $\pi \to Q^\pi$, then an even better policy $\pi^{+}$ could be derived as follows:
\[
\pi^{+} (s, g) = \argmax_{a} Q^{\pi}_g (s,a).
\label{eq:improvepolicy}
\]
This is the general theme of various iterative schemes: Acquire an approximation to $Q$, maximize it, and repeat.

Although we can never access $Q$ precisely, it can be estimated with samples from the environment.
Based on \cref{eq:returnt0}, the \emph{discounted return} accumulates rewards starting at some time index $t$:
\[
G_t = r_t + \gamma r_{t+1} + \gamma^2 r_{t+2} + \ldots = \sum_{k=0}^\infty \gamma^k r_{t+k}.
\label{eq:discountreturn}
\]
By averaging over trajectories, we find that 
\[
Q^{\pi}_g (s,a) = \EE_\pi \left[ G_0 \middle| s_0 = s, a_0 = a \right].
\label{eq:Qfunc}
\]
However, there is a rich structure we can exploit:
The discounted returns satisfy the recursion
\[
\begin{aligned}
G_t &= r_t + \gamma r_{t+1} + \gamma^2 r_{t+2} + \ldots \\
& = r_t + \gamma \left( r_{t+1} + \gamma r_{t+2} + \ldots \right) \\
& = r_t + \gamma G_{t+1},
\end{aligned}
\label{eq:tdreturn}
\]
which in turn implies (along with the Markov property) that $Q$ itself satisfies a tidy self-consistency relationship:
\[
Q^{\pi}_g (s,a) = r_g (s,a) + \gamma \EE_{s' \sim \pp{p}{s'}{s,a}, a' \sim \pp{\pi}{a'}{s', g}} \left[ Q^{\pi}_g (s', a') \right]. 
\label{eq:Qfixed}
\]

\Cref{eq:Qfixed} holds for any policy.
Naturally, define $Q^\star_g (s,a) = \underset{\pi}{\max}\ Q^{\pi}_g (s,a)$; indeed, if $Q^\star_g$ is available, then the optimal policy is obtained by
\[
\pi^\star (s, g) = \argmax_{a} Q^\star_g (s, a).
\label{eq:optpolicy}
\]
When we apply the recursion in \cref{eq:Qfixed}, the optimization is offset to the next time step:
\begin{equation}
\begin{aligned}
Q^\star_g(s,a) &= r_g(s,a) + \max_{\pi} \gamma \EE_{s' \sim \pp{p}{s'}{s,a}, a' \sim \pp{\pi}{a'}{s', g}} \left[ Q^{\pi}_g(s', a') \right] \\
&= r_g (s,a) + \gamma \EE_{s' \sim \pp{p}{s'}{s,a}} \left[ \max_{a' \in \action} \max_{\pi} Q^{\pi}_g (s',a') \right] \\
&= r_g(s,a) + \gamma \EE_{s' \sim \pp{p}{s'}{s,a}} \left[ \max_{a' \in \action} Q^\star_g (s',a') \right].	
\end{aligned}
\label{eq:bellmanQ}
\end{equation}
\Cref{eq:bellmanQ} is known as the \emph{Bellman optimality equation} \citep{sutton2018ReinforcementLearning, bertsekas1996neuro}.
Although we do not directly have access to $Q^{\pi}$, much less $Q^\star$, the beauty of \cref{eq:bellmanQ} is that it distills all the complexity of the original problem in \cref{eq:mdpobjective} into a one-step relation.
Essentially, the Bellman equation provides a principled theoretical target around which \ac{RL} algorithms are built.

The literature contains a vast number of algorithms  proposed  to solve \cref{eq:bellmanQ}.
We briefly mention two principles that pertain to future sections.

{\bf Learning from past experience.}\quad
Fix some policy $\pi$ and let it acquire experience in the form of \cref{eq:sarsa}.
Now let $\tilde{Q}$ be a tractable approximation of $Q^\star$. 
In the simplest case, $\tilde{Q}$ is a large table containing value estimates corresponding to a discrete set of state-action-goal pairs.
Importantly, $\tilde{Q}$ is some function we can evaluate at any $(s,a, g) \in \state \times \action \times \mathcal{G}$.
With our observed data $\{ s_t, a_t, r_t, s_{t+1}, \ldots \mid g \}$, $\tilde{Q}$ can be updated to encourage its predictions to satisfy \cref{eq:bellmanQ}:
\[
\tilde{Q}(s_t, a_t, g) \leftarrow \left( 1 - \alpha \right)\tilde{Q}(s_t, a_t, g) + \alpha \left( r_t + \gamma \max_{a' \in \action} \tilde{Q}(s_{t+1}, a', g) \right),
\label{eq:Qlearning}
\]
where $\alpha > 0$ is a step size.
Note that the policy $\pi$ that collected the data samples does not appear in this update equation, hence, \cref{eq:Qlearning} is emblematic of \emph{off-policy learning} methods.\footnote{\emph{On-policy} refers to the problem of learning $Q^\pi$.}
\Cref{eq:Qlearning} comes from $Q$-learning \citep{watkins1992Qlearning} and acts as inspiration for many deep \ac{RL} algorithms, popularized by \citet{mnih2013PlayingAtari} and \citet{silver2014DeterministicPolicy}.

More practically, consider a parameterized function approximator $Q_\phi$, such as a neural network, where $\phi$ represents a collection of scalar parameters---the trainable weights. For a given $\phi$,
$Q_\phi$ is an easy-to-evaluate function. We we want to choose $\phi$ so that $Q_\phi$ satisfies the Bellman optimality equation.
Given a tuple of data $(s, a, r, s', g)$ in dataset $\mathcal{D}$, define the target value:
\begin{equation}
q = r + \gamma \max_{a' \in \action} Q_\phi (s', a', g).
\label{eq:target}
\end{equation}
We can compare $Q_\phi (s,a,g)$ to $q$ and penalize $\phi$ for any mismatch.
In particular, we formulate the loss:
\[
\mathcal{L}(\phi) = \frac{1}{\abs{\mathcal{D}}} \sum_{(s, a, r, s', g) \in \mathcal{D}} \left( Q_\phi (s,a,g) - q \right)^2,
\label{eq:Qloss}
\]
where each $q$ is a tuple-dependent target defined in \cref{eq:target}, treated as training data independent of $\phi$ .
The parameters $\phi$ can then be updated using some form of gradient descent:
\[
\phi \leftarrow \phi - \alpha \grad \mathcal{L}(\phi).
\label{eq:updateQ}
\]
While these ideas give a template for learning complex policies from past experience, the underlying optimization procedure required to compute the targets in \cref{eq:target} can limit this approach in its nominal form.

{\bf Approximating the optimization process.}\quad
Based on \cref{eq:Qlearning}, a promising new policy can be designed as 
\[
\mu^{+} (s,g) = \argmax_{a} Q_\phi (s,a,g).
\label{eq:maxpolicy}
\]
However, the maximization can be expensive. Further, the optimization subproblem in \cref{eq:maxpolicy} must be solved not only during rollouts, but also in the update step in \cref{eq:updateQ} for each sample.
Therefore, introduce a ``nice'' parameterized policy $\mu_\theta$, and use a noisy version of $\mu_\theta$ for exploration:
\[
\pp{\pi}{a}{s,g} \sim \mathcal{N} \left( \mu_\theta (s,g), \Sigma \right).
\label{eq:noisepi}
\]

With both $\mu_\theta$ and $Q_\phi$ taking on some parameterization, they are referred to as the \emph{actor} and \emph{critic}, respectively \citep{konda1999ActorcriticAlgorithms, silver2014DeterministicPolicy}.
The idea is to use the policy to approximate the maximization operation in \cref{eq:maxpolicy}, and to use the critic to approximate the $Q$-learning target based on \cref{eq:target}.
That is,
\begin{equation}
\begin{aligned}
	q &= r + \gamma  Q_\phi (s', \mu_\theta(s', g), g)\\
	\phi &\leftarrow \phi - \alpha \grad_\phi \frac{1}{\abs{\mathcal{D}}} \sum_{(s, a, r, s', g) \in \mathcal{D}} \left( Q_\phi (s,a,g) - q \right)^2 \\
	\theta &\leftarrow \theta + \alpha \grad_\theta \frac{1}{\abs{\mathcal{D}}}  \sum_{(s, a, r, s', g) \in \mathcal{D}} Q_\phi (s, \mu_\theta (s, g), g).
\end{aligned}
\label{eq:dpgalg}
\end{equation}
The targets $q$ are now very simple to compute, only requiring function evaluation, rather than exact optimization:
\[
Q_\phi (s, \mu_\theta(s,g), g) \approx \max_{a \in \action} Q_\phi (s, a, g).
\label{eq:ac_value}
\]
This streamlines the rest of the updates, making it possible to iterate \cref{eq:dpgalg} over enormous datasets ($|{\mathcal D}|$) and very large parameter vectors ($\phi$ and $\theta$).

\subsection{Goal-conditioned learning}
\label{subsec:goalconditioned}

The \ac{RL} agent is tasked with achieving some goal efficiently.
However, the notion of efficiency is characterized by the reward function, which is often defined and fine-tuned by a user through various metrics and penalty terms \citep{andrychowicz2017HindsightExperience}.
Effectively designing a reward or stage cost is a common challenge in both \ac{RL} \citep{liu2022GoalConditionedReinforcement} and \ac{MPC}  \citep{forbes2015ModelPredictive}.
Ideally, one would only need to set a target goal $g$ and the agent would learn from a simple reward like this:
\begin{equation}
r_g (s,a) = 
\begin{cases}
	1 \quad \text{Goal is achieved} \\
	0 \quad \text{Otherwise}
\end{cases}
\label{eq:sparsereward}
\end{equation}
Naturally, a goal-conditioned policy produces actions $a \sim \pp{\pi}{a}{s,g}$ aimed at bringing the environment to goal $g$ and staying there.

A reward like \cref{eq:sparsereward} benefits from a great deal of flexibility.
Its minimal structure imposes no restrictions on the agent that affect \emph{how} it reaches its goal; rather, the agent only knows \emph{what} to achieve.
However, the signal produced by such a reward is extremely sparse.
Newly initialized policies are likely to accumulate a large cache of zeros.
Moreover, two different but suboptimal policies can fail in very different ways and yet receive the same feedback.

There are two paths forward:
\begin{enumerate}
	\item \textbf{Use dense rewards.}\quad Rewards, for example, of the form
		\[-r_g(s,a) = \left( s - g \right)\transpose M \left( s - g \right) + \left(\Delta a\right)\transpose R \left( \Delta a \right)\]
		provide a continuous signal to the agent that makes it easier to distinguish the utility of different actions. 
		While the meaning of the weight terms is straightforward, they are nuisance parameters that can dramatically affect how an ``optimal'' policy looks; see  \citet{forbes2015ModelPredictive} for a simple illustration.
	\item \textbf{Use hindsight.}\quad Learning through hindsight follows the premise that all trials---even ``unsuccessful'' ones---are informative \citep{andrychowicz2017HindsightExperience,eysenbach2020RewritingHistory}.
	Given a trajectory $\{ s_0, a_0, r_0, \ldots, s_T \mid g \}$
    deemed unsuccessful at achieving some goal $g$, the sequence of rewards would be all zeros. However, one thing is certain: Had $s_T$ been the goal, then the policy would have been successful. 
\end{enumerate}

\citet{andrychowicz2017HindsightExperience} first proposed \ac{HER}, that is, the  use of hindsight to learn goal-conditioned policies.
\Ac{HER} is not \iac{RL} algorithm, but rather a type of replay buffer that any off-policy algorithm can sample from.
For example, all the transition tuples $( s_t, a_t, r_t, s_{t+1}, g )$ in a goal-conditioned trajectory are kept:
\[
\{ s_0, a_0, r_0, s_1, a_1, r_1, \ldots s_T, a_T, r_T \mid g \}
\label{eq:HERbuffer}
\]
Additionally, define $s_T$ to be a fictitious goal.\footnote{We use the terminal state for simplicity. One may also sample future states from the trajectory.}
Then for each time step $i$ in \cref{eq:HERbuffer}, add the corresponding relabeled transition tuple to replay memory:
\[
( s_i, a_i, \underbrace{r_{s_T}(s_i, a_i)}_{\text{New reward}}, s_{i+1}, s_T ).
\label{eq:relabeled}
\]
The resulting replay memory contains both the original and hindsight-relabeled tuples, namely data from the original reward function $r_g$ and the relabeling reward function $r_{s_T}$; both rewards take the same functional form, only differing based on their target goal value, $g$ or $s_T$.
This results in additional ``excitation'' to the goal axis in the policy and value networks.
Even though \cref{eq:sparsereward} is likely to be sparse in the early stages of training, \cref{eq:relabeled} is guaranteed to contain at least one successful observation, namely, $r_{s_T}(s_T, a_T)=1$ because $s_T$ serves as a goal state in the dataset.
Over time, the agent learns a better correspondence between goals and actions, making it able to reliably reach the desired targets.

\section{Building a local value function through MPC}
\label{sec:localMPC}

\Ac{RL} hinges on the idea that an optimal policy can be discovered through a repeated cycle of exploration and improvement.
The subtext of this paradigm is that such a policy should be learned from scratch.
However, many control applications entail some prior physical understanding of the system, opening up opportunities to warm start the policy search \citep{venkatasubramanian2019PromiseArtificial}.\footnote{In this section, we do \emph{not} refer to model-based \ac{RL} wherein a dynamical model is learned or made available to aid in the training of the \ac{RL} agent with otherwise model-free algorithms \citep{jafferjee2020HallucinatingValue,janner2019WhenTrust}.}
Here, we provide an outline of nominal \ac{MPC}.
That is, an exact model is available so as to emphasize the complete opposite of the \ac{RL} approach of the previous section.

\Ac{MPC} is the most successful advanced control method \citep{qin2003Surveyindustrial,lee2011Modelpredictive,schwenzer2021Reviewmodel}.
It is ``safe'', modular, and interpretable:
\begin{itemize}
	\item \textbf{Safe.}\footnote{\Ac{MPC} is \emph{not} a magic bullet. Our point is that the \ac{MPC} literature provides a theoretical blueprint for formulating safe policies comprising technical conditions regarding stability, robustness, optimality, and constraint handling \citep{borrelli2017predictive}.}\quad Model knowledge and other constraints help compose an objective whose optimal solution leads to safe and stable operations.
	\item \textbf{Modular.}\quad Individual components of the controller can in principle be modified on the fly to reflect new knowledge or objectives.
	\item \textbf{Interpretable.}\quad The combination of constraints and modularity makes \ac{MPC} an intuitive approach for control (notwithstanding the underlying technical requirements).
\end{itemize}

\subsection{An analytical foundation for MPC}

Rather than using samples from the environment to learn a value function, this section focuses on constructing a value function.
This is done by combining a dynamic model and a cost function.
We begin with the \ac{LQR} problem:\footnote{We pivot to a \emph{minimization} problem, versus a \emph{maximization} problem in \ac{RL}, purely due to convention.}
\begin{equation}
\begin{aligned}
    &\underset{\mu(\cdot)}{\text{minimize}} && \sum_{t=0}^{\infty} \gamma^{t} \left( x_t\transpose M x_{t} + u_{t}\transpose R u_t \right) \\
    &\text{subject to } && u_t = \mu(x_t)\\
    &	&& x_{t+1} = A x_t + B u_t.
\end{aligned}
\label{eq:LQRobjective}
\end{equation}
Here the state $x$ and control $u$ take values in $\mathbb{R}^n$ and $\mathbb{R}^m$, respectively; the dynamics involve known matrices $A$ and $B$ of compatible dimension starting at an arbitrary initial state $x_0 = \bar{x}$; and user-selected symmetric positive-definite matrices $M$ and $R$ determine the cost of deviations from the nominal trajectory where both $x$ and $u$ are constant at the origin.
This is the simplest nontrivial case of the global objective in \cref{eq:mdpobjective} for which there is an analytical solution \citep{bertsekas2022lessons}.
This additional structure makes the new problem in \cref{eq:LQRobjective} seem more palpable than the original:
It considers a linear, time-invariant environment and a global objective that can be characterized by a quadratic cost around the origin.
Moreover, the optimization is over deterministic policies $\mu$.

The optimal solution to the \ac{LQR} problem is a static linear controller $\mu(x_t) = -K x_t$.
A key step in the solution is the use of the Bellman equation in tandem with a quadratic value function $V^\star (x) = x\transpose P x$ (see \cref{app:lqr} for more details):
\[
x\transpose P x = \min_{u} \left\{ x \transpose M x + u\transpose R u + \gamma \left(A x + B u\right)\transpose P \left(A x + B u\right) \right\},
\]
wherein solving for $u$ leads to an explicit formula for $K$.
This is a powerful result.
The \ac{LQR} problem not only yields a quadratic \emph{global} value function, but its simple structure lends itself to a tractable solution.
This means we are now equipped with a formula that takes system and cost parameters and maps them to an optimal set of controller parameters.\footnote{Here ``optimal'' must be understood in the narrow context of \cref{eq:LQRobjective}; generally, arranging for the closed-loop dynamics to meet practical performance criteria requires careful choice of the matrices $M$ and $R$. Requiring both to be positive definite suffices to guarantee closed-loop stability of the optimal controller, a key safety requirement in practice.}

\subsection{MPC as an implicit control law}

In light of the \ac{LQR} objective in \cref{eq:LQRobjective}, it is natural to wonder about the possibility of additional constraints:
\begin{equation}
\begin{aligned}
    &\underset{\mu(\cdot)}{\text{minimize}} && J(\mu) = \sum_{t=0}^{\infty} \gamma^{t}\left( x_t\transpose M x_{t} + u_{t}\transpose R u_t \right)\\
    &\text{subject to } && u_t = \mu(x_t)\\
    & && x_{t+1} = A x_t + B u_t\\
    & && x_t \in \mathcal{X}, u_t \in \mathcal{U}.
\end{aligned}
\label{eq:feedbackobjective}
\end{equation}
This builds on \cref{eq:LQRobjective} by asserting that system behavior requirements are captured by state-input constraint sets $\mathcal{X} \times \mathcal{U}$, often box constraints.

A controller resulting from the constrained problem in \cref{eq:feedbackobjective} is inherently nonlinear.
Indeed, control actions are state-dependent, as they account for proximity to the constraints.
This contrasts with the \ac{LQR} solution, which applies the same operation to the state no matter what.
Thus, the \ac{LQR} solution is not the best solution to the constrained problem, as it may only remain feasible inside a ``small'' portion of the state space \citep{borrelli2017predictive}.
In \cref{eq:feedbackobjective}, one could consider a parameterized class of policies (i.e., state feedback controllers) $\mu_\theta$ and proceed in a similar fashion to the \ac{RL} approach.
The result would be an explicit mapping $\mu_\theta: \state \to \action$ acting on the true environment.
However, this mapping introduces a degree of separation from the prior knowledge embedded in \cref{eq:feedbackobjective}, such as the system dynamics and cost structure.
In contrast, \ac{MPC} offers an implicit formulation aimed at retaining the design elements appearing in \cref{eq:feedbackobjective}. 
We outline two core features of the \ac{MPC} approach.

{\bf Preserving prior knowledge and requirements.}\quad
\Cref{eq:feedbackobjective} can equivalently be cast in terms of a sequence of inputs $u_0, u_1, u_1, \ldots$:
\begin{equation}
\begin{aligned}
    &\underset{u_0, u_1, u_2, \ldots}{\text{minimize}} && \sum_{t=0}^{\infty} \gamma^{t} \left( x_t\transpose M x_{t} + u_{t}\transpose R u_t \right) \\
    &\text{subject to } && x_{t+1} = A x_t + B u_t \\
    & && x_t \in \mathcal{X}, u_t \in \mathcal{U}.
\end{aligned}
\label{eq:desiredMPCobj}
\end{equation}
However, this problem contains an infinite number of decision variables.
A pragmatic idea is to formulate a hybrid between \cref{eq:desiredMPCobj,eq:LQRobjective}.
Consider the new objective, defined at some state $s$:
\begin{equation}
\begin{aligned}
    &\underset{u_0, u_1, \ldots, u_{N_c-1}}{\text{minimize}} && \sum_{t=0}^{N-1} \gamma^t \left( x_t\transpose M x_{t} + u_{t}\transpose R u_t \right) + \gamma^{N} x_{N}\transpose P x_N \\
    &\text{subject to } && x_0 = s \\
    & && x_{t+1} = A x_t + B u_t \\
    & && x_t \in \mathcal{X}, u_t \in \mathcal{U} \\
    & && u_t = -K x_t,\quad N_c \leq t \leq N-1.
\end{aligned}
\label{eq:nominalMPCobj}
\end{equation}
This new problem considers a finite number of decision variables, enabling reasonable command over constraints and system knowledge, while embedding infinite-horizon behavior cached in the \ac{LQR} value function \citep{lee2011Modelpredictive}.

{\bf Building a local value function.}\quad
The standard \ac{MPC} algorithm implements a \emph{receding horizon} strategy:
After solving \cref{eq:nominalMPCobj} at some state $s$ for optimal inputs $u_{0}^{\star}, \ldots, u_{N_{c}-1}^{\star}$, the action $ a = u_{0}^{\star}$ is applied to the true system.
The system transitions to some next state $s'$, at which point the problem in \cref{eq:nominalMPCobj} is reinitialized and solved again.

The use of $K$ as a ``fictitious'' controller in \cref{eq:nominalMPCobj} and $P$ as a terminal cost enable feasibility and stability guarantees \citep{borrelli2017predictive}.
Without them, perhaps by truncating the objective, the repeated application of solutions to \cref{eq:nominalMPCobj} is not guaranteed to always be feasible, much less stable.
Essentially, without incorporating infinite-horizon knowledge into the problem, anything beyond $N_c$ steps comes as a ``surprise'' to the controller.

All taken together, the receding horizon idea in tandem with the structure in \cref{eq:nominalMPCobj} represent an implicit, \emph{local} value function approximation \citep{mayne2000Constrainedmodel}.
Costs and actions are computed online as new state information is made available.
Crucially, the practical and theoretical success of \ac{MPC} is driven by this interplay between a global \ac{LQR} value function and local replanning.
The global \ac{LQR} solution uses principles of dynamic programming to cache all the planning into an explicit policy.
In turn, this alleviates the intractability of infinite-horizon planning as in \cref{eq:desiredMPCobj}.

\section{A brief survey of the RL-MPC interface}
\label{sec:related}

We now discuss related studies at the intersection of \ac{RL} and \ac{MPC}.
Value functions are fundamental to \ac{MPC} theory to derive stability and recursive feasibility conditions \citep{mayne2000Constrainedmodel, borrelli2017predictive, abdufattokhov2024LearningLyapunov}, which is not the focus of this paper.
Moreover, we do not survey learning-based approaches to MPC \citep{hewing2020LearningBasedModel, mesbah2022FusionMachine} or design of safety filters \citep{wabersich2021Predictivesafety, bejarano2024SafetyFiltering, hosseinionari2024IntegrationModel}, where learning generally plays a supporting role to enhance safety and performance of MPC.   
Instead, we focus on learning-based control strategies that take advantage of conceptual similarities between \ac{RL} and \ac{MPC}.
In particular, our brief survey adopts a value-centric perspective, reflecting the core framework of this paper.
More detailed surveys on the \ac{RL}-\ac{MPC} interface can be found in \citet{reiter2025SynthesisModel}, \citet{banker2025ModelfreeReinforcement}, and \citet{banker2025LocalGlobalLearning}.

{\bf Value function-augmented \ac{MPC}.}\quad 
Our approach falls into this category because we use \iac{MPC} agent to design actions using a learned value function.
However, the basic idea of \iac{RL}-based value function-augmented \ac{MPC} law is not new. 
This approach is based on dynamic programming, but made practical through \ac{RL} techniques.\footnote{Some of the referenced works use terms like \emph{\ac{ADP}} or \emph{neuro-dynamic programming}. We use \emph{\acl{RL}} for simplicity.}
Foundational works by \citet{bertsekas1996neuro} provide a rigorous treatment of \ac{RL}, while \citet{bertsekas2022lessons} gives a more recent account with emphasis on value function approximation and \ac{MPC}.

The key idea behind value function-augmented \ac{MPC} approaches is to consider \iac{MPC} objective of the form
\begin{equation}
	\sum_{t=0}^{N-1} \gamma^t \hat{r}(x_t, u_t) + \gamma^{N} V(x_N),
\end{equation}
where the state-action trajectory is produced by a dynamic model, possibly subject to constraints and over uncertain scenarios.
$V$ is learned through \ac{RL} techniques such as the nominal scheme in \cref{eq:dpgalg}.
In particular, we have
\begin{equation}
	V(x_N) = Q_\phi (x_N, \mu_\theta(x_N)) \approx \max_{u \in \action} Q_\phi (x_N, u),
\end{equation}
as in \cref{eq:ac_value}, ignoring goals for simplicity.
Meanwhile, the $N$-step predictions are carried out by a dynamic model and guided by the reward $\hat{r}$.
Crucially, these components are acquired independently of the terminal value function $V$.
The model can be learned under some prediction-based objective, or given through prior knowledge; similarly, $\hat{r}$ can be an optimization-friendly approximation of the true reward.
Simply put, the distinguishing characteristic of value-augmented \ac{MPC} is its \emph{modularity}.

Early works by \citet{lee2001NeurodynamicProgramming, lee2004SimulationbasedLearning} demonstrated the utility of embedding a learned value function into \ac{MPC} for process control applications.
\Citet{zhong2013Valuefunction} apply similar ideas in the context of classic control problems with an emphasis on data collection and value function parameterizations.
Similarly, the works of \citet{lowrey2019PlanOnline} consider \ac{MPC} as a trajectory optimizer that can aid in value function estimation, but with emphases on exploration.
So far, these works assume \ac{MPC} uses a locally optimal value function, meaning the cost and internal model accurately represent the true objective and environment. 
Nonetheless, a key benefit of value function-augmented \ac{MPC} via \ac{RL} is the ability to effectively shrink the planning horizon. Instead, a significant amount of planning and uncertainty can be cached into the value function representation, which lends itself nicely to stochastic systems.

\Citet{farshidian2019DeepValue} consider the case where an external, possibly sparse, reward signal is used to update the stage cost and value function in \ac{MPC}, but still assume an accurate model.
\Citet{arroyo2022ReinforcedModel} train \iac{RL} agent offline in simulation with an identified model, then deploy a value function-augmented \ac{MPC} scheme on the true, more complex system.
However, the agent remains static in the online phase, not taking into account information from the environment.
On the other hand, \citet{bhardwaj2020BlendingMPC} devise a time-weighted averaging strategy that blends together the \ac{MPC} and \iac{RL}-learned value estimate, taking advantage of prior information while enabling feedback from the true environment.
\citet{hansen2024TDMPC2Scalable} develop a complete learning pipeline in which the model, reward, and value function are all learned and used to construct an online \ac{MPC} agent.

{\bf \ac{MPC} as a function approximator.}\quad
Another line of work takes the view that \ac{MPC}---its model, stage cost, constraints, and terminal value function---represents a set of parameters that can be steered towards closed-loop optimality under a single objective.
Although value function-augmented \ac{MPC} takes a modular view of learning-based \ac{MPC}, this alternative formulation presents an \emph{all-in-one} perspective.
While we emphasize \ac{RL}-based methods, this perspective can also be generalized to include other policy search paradigms \citep{hu2023TheoreticalFoundation, banker2025ModelfreeReinforcement}.

This all-in-one view of learning-based \ac{MPC} is compatible with both policy-based and value-based \ac{RL} methods.
The key idea is to replace traditional function approximations, such as \aclp{DNN}, in \iac{RL} pipeline with \ac{MPC}.
For example, \cref{eq:nominalMPCobj} can be viewed as a $Q$-function parameterization, which we will call $Q_\phi^{\text{MPC}}$.
Conceptually, typical \ac{RL} update steps proceed as usual, where the goal is to minimize the loss in \cref{eq:Qloss} via gradient steps in \cref{eq:updateQ}.
A similar idea can be applied to the policy parameterization, where we focus on the minimizer of \cref{eq:nominalMPCobj}, and take $\mu_\theta^{\text{MPC}}$ to be a parameterized \ac{MPC} policy.
In either case, all the \ac{MPC} parameters are free to update towards improved reward.
One may also choose to update only a subset of the \ac{MPC} parameters in this manner; doing so is still distinct from the value function-augmented approach because the \ac{MPC} structure is embedded in the \ac{RL} pipeline, such as \cref{eq:dpgalg}.

A common approach to learning \ac{MPC} policies is to differentiate through the \ac{MPC} action with respect to its parameters.
\Citet{amos2019DifferentiableMPC} propose this idea, but apply it for imitation learning tasks. 
In a similar vein, \citet{tamar2017learning} iteratively refine the \ac{MPC} cost based on offline replanning.
\Citet{gros2020DataDrivenEconomica, gros2022LearningMPC} further develop this line of work with an emphasis on safety and stability under \ac{RL}-based updates to the \ac{MPC} parameters.
In the context of deep \ac{RL}, \citet{romero2024ActorCriticModel} propose an actor-critic setup in which the actor feeds cost coefficients to a differentiable \ac{MPC} module.
See \citet{reiter2025SynthesisModel} and \citet{banker2025ModelfreeReinforcement} for an extensive discussion of \ac{MPC}-based function approximators and their respective formulations.

Broadly speaking, these approaches place less trust in prior system knowledge than value function-augmented approaches and, instead, aim to find the best model---dynamics, constraints, and cost---for control, inspired by the notion of identification for control \citep{gevers2005identification}.
This all-in-one approach makes \ac{MPC} inextricably linked to the underlying \ac{RL} algorithm.
This means \iac{MPC} parameterization exists explicitly inside the nominal \ac{RL} update equation in \cref{eq:dpgalg}, targeting \emph{global} performance in the Bellman sense, while serving as a powerful inductive bias for modeling \emph{local} values (or actions) for optimization-based decision-making~\citep{banker2025LocalGlobalLearning}.
It is worth noting that this requires computing either $\nabla_\phi Q_{\phi}^{\text{MPC}}$ (or $\nabla_\theta \mu_{\theta}^{\text{MPC}}$).
These are nontrivial computations and the subject of ongoing research and software development \cite{reiter2025SynthesisModel}.
Meanwhile, value function-augmented schemes allow for more algorithmic separation, allowing the value function to be trained in a deep \ac{RL} pipeline, possibly offline based on prior system knowledge, and ported to the online \ac{MPC} agent.
However, despite this separation in the learning pipeline, \ac{RL} and \ac{MPC} are still intimately connected together through their roots in the dynamic programming formalism.
While value function-augmented \ac{MPC} may not suffer from the abovementioned gradient calculations, its online optimization may be expensive; we revisit this point in \cref{app:LQ-GC}.

\section{Robust goal-conditioned control policies}
\label{sec:rlmpc}

\begin{figure}
	\includegraphics[width=0.9\textwidth]{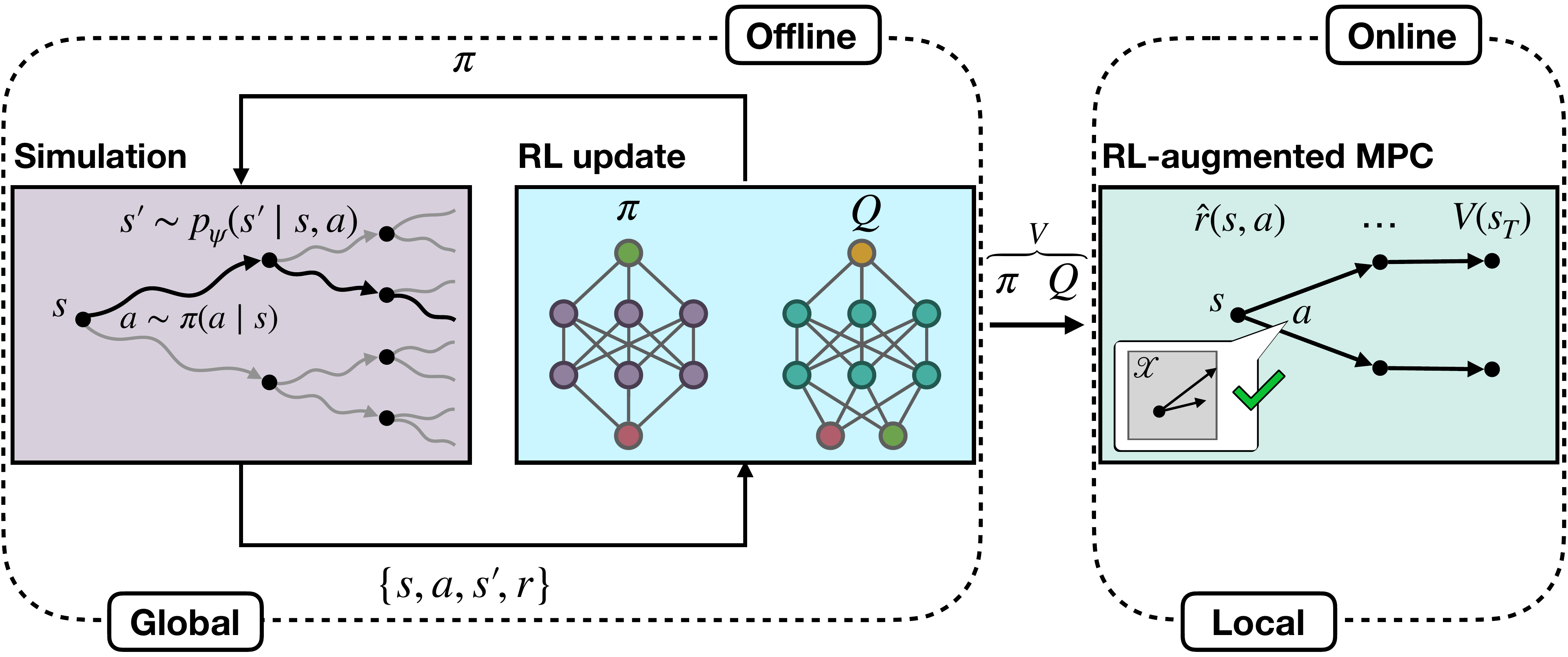}
	\caption{An actor-critic agent interacts with a branching simulation environment offline to learn a robust global value function. The critic is used in the usual fashion to inform parameter updates, but also to construct a robust local \ac{MPC} agent for online control of the ``true'' system.}
	\label{fig:concept}
\end{figure}

This section builds on the local-global interface through robust \ac{MPC} and goal-conditioned \ac{RL}. 
We first extend the discussion of nominal \ac{MPC} to robust \ac{MPC}.
This then inspires a robust training scheme for goal-conditioned \ac{RL}.
Finally, we show how to combine these agents such that the \ac{RL} policy benefits from replanning and constraint handling, while the \ac{MPC} policy benefits from high-level goal-conditioned objectives. 
See \cref{fig:concept} for an illustration of the proposed framework.

\begin{figure}
	\includegraphics[width=3.3in]{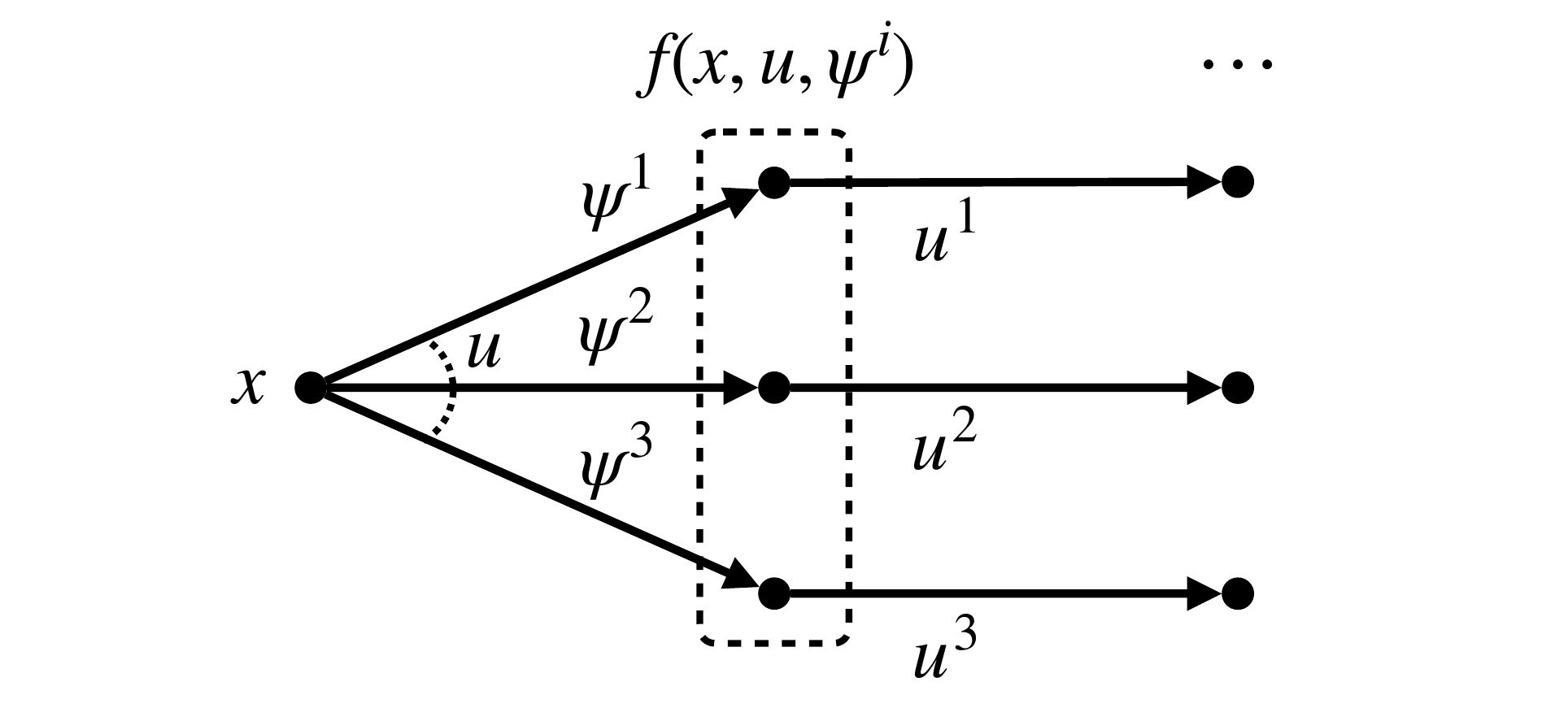}
	\caption{A scenario tree branches at some state $x$, applying the same control action $u$ for each of the three cases in the uncertainty set $\{\psi^1, \psi^2, \psi^3\}$. Three successive states are computed using the model $f$, after which each scenario remains constant. However, it is possible to keep branching at each node.}
	\label{fig:scenario}
\end{figure}

Our framework is model-based in nature.
Specifically, we assume an \emph{uncertain} dynamic model of the environment is available: We have some prior knowledge, but not enough to support a perfect representation of the underlying dynamics.
Robustness is incorporated into our framework through the uncertain system description:
\begin{enumerate}
	\item \textbf{Robustness of the online agent.}\quad We use a robust scenario-based \ac{MPC} agent, which incorporates a distribution of system uncertainties into its predictions. Specifically, the \ac{MPC} agent constructs a scenario tree, illustrated in \cref{fig:scenario}, to tabulate costs and account for constraints over different situations.  
	\item \textbf{Robustness of the offline agent.}\quad We formulate a scenario-based value function based on the distribution of system uncertainties. This leads to a robust Bellman equation, which serves as a target for the \ac{RL} agent to learn simply through a branching process during offline rollouts; see the left-hand portion of \cref{fig:concept}. General off-policy actor-critic algorithms are applicable for this portion of the framework \citep{konda1999ActorcriticAlgorithms}.
\end{enumerate}
Essentially, this scenario-based approach to robustness aligns the robust \ac{RL}-learned value function with the short-term, uncertain \ac{MPC} predictions.

Although a model is available, we target complicated objectives where a straightforward implementation of \ac{MPC} may not be suitable.
Consequently, we leverage model-free \ac{RL} techniques to directly learn the optimal value function from offline exploration.
In particular, the \ac{RL} agent is trained in a robust, goal-conditioned manner. 
After training such \iac{RL} agent offline, \iac{MPC} agent generates actions using short-term predictions and the \ac{RL} value function as a terminal cost.
Specifically, the \ac{MPC} agent makes these predictions subject to constraints and system uncertainty.
Together, this combination of \ac{RL} and \ac{MPC} produces a robust and safe goal-conditioned policy.

\subsection{Scenario-based MPC}
\label{subsec:multiMPC}

The \ac{MPC} formulation given in \cref{eq:nominalMPCobj} is often referred to as \emph{nominal MPC} wherein one assumes the system model reflects the true dynamics being controlled.
Yet, the basic idea of continually replanning endows the basic \ac{MPC} structure with some inherent robustness to plant-model mismatch.
Nonetheless, the risk of violating constraints when deploying \iac{MPC} scheme should not be overlooked.
Our proposed framework employs a scenario-based approach to robustness \citep{lucia2013MultistageNonlinear}.
We note other varieties of robust and stochastic \ac{MPC} \cite{bemporad2007robust,mesbah2016stochastic} may be used during the online phase of the proposed framework. 
For example, popular robust MPC alternatives include tube-based \ac{MPC} \citep{mayne2005robust} and min-max \ac{MPC} \citep{campo1987RobustModel}.
Scenario-based \ac{MPC} does not require a precomputed ancillary controller, as in tube-based \ac{MPC}, nor does it generally lead to a very conservative solution, as in min-max \ac{MPC}.
Instead, the scenario-based approach has its backbone in dynamic programming, making it a unified target for approximate solutions through both \ac{MPC} and \ac{RL} \citep{delapenad2005StochasticProgramming,bernardini2009Scenariobasedmodel}.

Scenario-based \ac{MPC} considers a scenario tree in its planning to help cope with uncertainty \citep{lucia2013MultistageNonlinear}. 
A scenario is essentially a realization of the system model under some uncertainty specification.
System uncertainty is general under our framework, but some possible sources include structural model uncertainty, model parameter uncertainty, or time-varying components \citep{paulson2018NonlinearModel}.
Each scenario is subject to the same constraints, which means actions that are otherwise reasonable under nominal MPC may get pruned from consideration.
This strategy results in more robust actions.

Mathematically, consider a general system model $f$ whose successive state $x'$ evolves as follows:
\begin{equation}
	x' = f(x, u, \psi).
\label{eq:paramsys}
\end{equation}
In addition to states $x$ and control actions $u$, $f$ also takes in scenarios $\psi$.
We assume $f$ is given through prior physical understanding of the process, but $\psi$ represents system uncertainty due to structural and parametric model uncertainty or exogenous disturbances.
Consider $N_s$ scenarios, each of which is a realization of the uncertainties $\psi$ in a system model $f$, branching from the start state then remaining constant; \cref{fig:scenario} illustrates the basic concept. 
It is possible to branch out the uncertainty scenarios at each time step, but this is discouraged due to the exponential growth in scenarios.

Controlling the growth in the number of scenarios is a practical innovation of scenario-based \ac{MPC} \citep{lucia2013MultistageNonlinear}, which considers the following objective at some initial state $s$:\footnote{The form given by \citet{lucia2013MultistageNonlinear} includes a \emph{robustness horizon} parameter, which controls how many time steps branch out in the prediction horizon. We present the case where the robustness horizon is $1$.}
\begin{equation}
\begin{aligned}
    &\underset{\substack{u_0^1, u_1^1, \ldots, u_{N-1}^1\\\vdots\\ u_0^{N_s}, u_1^{N_s}, \ldots, u_{N-1}^{N_s}}}{\text{minimize}} && \frac{1}{N_s} \sum_{i=1}^{N_s} J\left( x_0^i,\ldots x_{N}^i, u_0^i, \ldots, u_{N-1}^i \right) \\
    &\text{subject to } && x_0^i = s \\
    & && x_{t+1}^i = f(x_t^i, u_t^i, \psi^i) \\
    & && u_{0}^i = u_{0}^j \\
    & && x_t^i \in \mathcal{X}, u_t^i \in \mathcal{U} \\
    & && x_{N}^i \in \mathcal{X}_{\text{terminal}}
\end{aligned}
\label{eq:multiMPCobj}
\end{equation}
where $J$ is an $N$-step cost function:
\[
J(x_0,\ldots x_{N}, u_0, \ldots, u_{N-1}) = \sum_{t=0}^{N-1} l(x_t, u_t) + m(x_N).
\]
This formulation considers a general \emph{stage cost} $l$ and \emph{terminal cost} $m$.
In the context of \cref{eq:nominalMPCobj}, the stage cost is quadratic and the terminal cost is the \ac{LQR} value function.
Note the constraint $u^i_0 = u^j_0$ ensures that the agent selects actions only according to current information $s$.
Successfully solving \cref{eq:multiMPCobj} provides a certificate that the optimal solution satisfies the constraints even under the worst-case scenario.
This provides some extra assurance that the endorsed action will keep the true system operating safely.

\subsection{Offline MDP based on uncertain knowledge}
\label{subsec:offlinemdp}

Beyond scenario-based \ac{MPC}, we utilize the idea of a scenario tree to formulate a branching \ac{MDP}, which can be used to train a robust \ac{RL} agent.
This is in contrast to other approaches to robustness in \ac{RL}.
A min-max formulation is a common strategy for training robust, although conservative, policies \citep{zouitine2024Solvingrobust, nilim2003RobustnessMarkov}.
Another algorithmic approach is to train conservative agents with respect to static, \emph{offline} datasets, leading to robustness in \emph{online} performance \citep{kumar2020ConservativeQlearning}.
Other approaches focus on imposing structural constraints on the policy architecture based on \acp{IQC} to achieve robustness \citep{jin2020StabilitycertifiedReinforcement, revay2021RecurrentEquilibrium}.
The proposed scenario-based approach is both simple and congruous with the overarching \ac{MDP} framework, as discussed next.

In the context of \iac{MDP}, structural knowledge of the model $f$ can be combined with the uncertainty in $\psi$ to formulate an environment.
That is,
\begin{equation}
\begin{split}
	\psi &\sim p(\psi)	\\
	s' &\sim \pp{p_\psi}{s'}{s,a}
\label{eq:offlineMDP}
\end{split}
\end{equation}
where $\pp{p_\psi}{s'}{s,a} = \delta\left( s' - f(x,u,\psi) \mid s = x, a = u \right)$ is the Dirac delta function conditioned on the current state-action pair.
The dynamics in \cref{eq:offlineMDP} define the transitions for an offline simulation environment.
Finally, a reward, such as in \cref{eq:sparsereward}, completes the \ac{MDP}.

\textbf{A robust value function.}\quad
We consider a set of $N_s$ possible realizations of system uncertainty:
\[
\{ \psi^0, \ldots, \psi^{N_s-1} \}.
\label{eq:paramset}
\]
No preference is given to any one of them, meaning they are uniformly distributed.
In the context of \cref{eq:offlineMDP} and the Bellman equation in \cref{eq:Qfixed}, we have the following relationship:
\begin{align}
\begin{split}
	Q^{\pi}_g (s,a) &= r_g (s,a) + \gamma \EE_{\psi \sim p(\psi), s' \sim \pp{p_\psi}{s'}{s,a}, a' \sim \pp{\pi} {a'}{s',g}} \left[ Q^{\pi}_g (s', a') \right] \\
	&= r_g (s,a) + \gamma \frac{1}{N_s} \sum_{i = 0}^{N_s-1} \EE_{s' \sim \pp{p_{\psi^{i}}}{s'}{s,a}, a' \sim \pp{\pi}{a'}{s',g}}\left[ Q^{\pi}_g (s', a') \right].
\label{eq:robustQ}
\end{split}
\end{align}

This theoretical target is in competition with some straightforward options regarding robustness:
\begin{enumerate}
	\item One could opt for a single uncertainty instance and hope that the resulting policy generalizes well to other scenarios.
	\item Going further, one could create multiple scenarios in parallel, sharing the same policy, then pool together the respective value functions $Q_{0},\ldots, Q_{N_s-1}$ through averaging $\frac{1}{N_s} \sum_{i = 0}^{N_s-1} Q_{i}$. This forms an approximation to $Q^{\pi}_g$ in \cref{eq:robustQ}, but ultimately does not result in a value function itself for the \ac{MDP} in \cref{eq:offlineMDP}. 
\end{enumerate}

Instead, in the spirit of option (1), the branching \ac{MDP} is only a single (but specialized) environment, but experience from all scenarios informs the value estimation, like option (2).
However, unlike these options, a policy satisfying \cref{eq:robustQ} directly incorporates uncertainty into the decision-making process in a state-dependent fashion. 
This means it has to operate with enough margin to elevate the next-step return across several scenarios.

\subsection{Offline training and online deployment}

The simulated \ac{MDP} in \cref{eq:offlineMDP} enables an agent to learn a goal-conditioned policy under uncertainty.
For our general formulation, any off-policy actor-critic algorithm can be used wherein a policy $\pi$ is learned alongside a value function $Q$ \citep{konda1999ActorcriticAlgorithms}.
Both are represented by \aclp{DNN}.
Briefly, the policy network $\pi$ is used in the simulation environment to enable ``fast'' decision-making and streamlined implementation.
We then deploy the critic $Q$ on the ``true'' system, where \iac{MPC} agent designs actions subject to constraints and uncertainty intervals.
A concept diagram summarizing this section is shown in \cref{fig:concept}.

{\bf Actor-critic training.}\quad
Based on \cref{eq:robustQ}, \iac{RL} algorithm seeks to learn $\pi$ and $Q$ such that:
\[
Q^{\pi}_g (s,a) \approx r_g (s,a) + \gamma \frac{1}{N_s} \sum_{i = 0}^{N_s-1} \EE_{s' \sim \pp{p_{\psi^{i}}}{s'}{s,a}}\left[\max_{a'} Q^{\pi}_g (s', a') \right].
\label{eq:approxQ}
\]
First, a dynamic model class is created based on \cref{eq:paramsys}.
This model structure is the basis for the environment in \cref{eq:offlineMDP}.
Such an environment has two key elements:
\begin{enumerate}
	\item \textbf{Branched rollouts.}\quad Sampling from the scenario set in \cref{eq:paramset} at each time step to create branched rollouts. Note that the scenario set used for offline \ac{RL} training may be larger than the one in scenario-based \ac{MPC} because the learned value function in \cref{eq:approxQ} does not perform explicit planning upon deployment.
	\item \textbf{Goal-augmented state.}\quad For goal-conditioned learning, the state definition used in the environment contains the goal itself\footnote{Equivalently, we use the error signal $g-s$, rather than the goal, as input to the actor-critic networks.}, the observed state, in the spirit of \cref{eq:paramsys}, as well as the ``achieved goal.'' The achieved goal could be the state itself, or some transformed version of the state, for instance, if the goal is an output value rather than a state value. All the information is necessary in order to implement the \ac{HER} strategy from a replay buffer.
\end{enumerate}
With the environment ready, an off-the-shelf off-policy, deep \ac{RL} algorithm can be deployed aimed at learning $\pi$ and $Q$ in \cref{eq:approxQ}.
The correct state formulation allows for the \ac{HER} strategy to be used to relabel training samples drawn from the replay buffer and used for updating the actor-critic weights.

{\bf Critic-informed \ac{MPC} deployment.}\quad 
A key innovation of deep \ac{RL} algorithms is the ability to train complex policies while avoiding exact optimization.
Specifically, the policy $\pi$ is trained to optimize $Q$, but only approximately, as discussed around \cref{eq:noisepi,eq:dpgalg}.
This amounts to using $Q$ as a loss function in training, and then the fast-to-evaluate $\pi$ for decision-making.

While the agent explores and learns in the offline MDP in \cref{eq:offlineMDP} through the policy $\pi$, the corresponding value approximation $Q$ is used in conjunction with \iac{MPC} agent to create a refined policy for online deployment.
Like the \ac{RL} policy $\pi$, this new, refined policy is also goal-conditioned.
It uses a Gaussian-shaped reward $\hat{r}_g$ with a fixed variance  $\sigma^2$:
\begin{equation}
	\hat{r}_g (s,a) = e^{-\frac{\norm{g-s}^2}{2 \sigma^2}} \approx 
	\begin{cases}
	1 & \text{Goal $g$ is achieved}\\
	0 & \text{Otherwise}.	
	\end{cases}
\label{eq:goalreward}
\end{equation}
The right-hand side represents an idealized reward signal.
However, it is non-smooth and the condition ``goal is achieved'' can be difficult to characterize; for instance, a typical heuristic is to assign $1$ if the distance to the goal is within some tolerance \cite{liu2022GoalConditionedReinforcement, andrychowicz2017HindsightExperience}.
In contrast, the functional definition on the left-hand side is a smooth approximation of the right-hand side with the additional benefit that its maximum is precisely at $s=g$.
Thus, the reward $\hat{r}_g$ aligns the short-term costs with the terminal, goal-conditioned value function $Q$.
Our experimental evaluation examines the variance parameter; a small variance is not necessary, as the short-term predictions are primarily concerned with the constraints, while the terminal value function provides more fine-grained guidance toward the goal.
Now, define the unified \ac{RL} and \ac{MPC} policy based on the following objective:
\begin{equation}
\begin{aligned}
&\underset{\substack{u_0^1, u_1^1, \ldots, u_{N-1}^1\\\vdots\\ u_0^{\hat{N}_s}, u_1^{\hat{N}_s}, \ldots, u_{N-1}^{\hat{N}_s}}}{\text{minimize}} && \frac{1}{\hat{N}_s} \sum_{i=1}^{\hat{N}_s} \sum_{t=0}^{N-1} \left[ \norm{\epsilon_t^i}_1 - e^{-\frac{\norm{g - x_{t}^{i}}^2}{2 \sigma^2}} \right] - V^{\pi}_g (x_{N}^{i}) \\
    &\text{subject to } && x_0^i = s \\
    & && x_{t+1}^i = f(x_t^i, u_t^i, \hat{\psi}^i) \\
    & && u_{0}^i = u_{0}^j \\
    & && x_t^i - \epsilon_t^i \in \mathcal{X}, \epsilon_t^i \in \mathcal{E}, u_t^i \in \mathcal{U},\\
\end{aligned}
\label{eq:QMPC}
\end{equation}
where $\hat{\psi}^i$ represents scenarios from a restricted subset of those used for \ac{RL} training and $\hat{N}_s$ is the corresponding number of scenarios.
As with any \ac{MPC}-based policy, only the first action in \cref{eq:QMPC} is deployed.
There are several pieces to unpack:
\begin{itemize}
	\item \textbf{Terminal cost.}\quad We use the learned value function $V^{\pi}_g (s) = Q^{\pi}_g (s, \mu_{\text{actor}}(s))$ as a terminal cost, where $\mu_{\text{actor}}$ is the mean of the policy $\pi$. Rather than implementing the \ac{RL} actions directly, the value function informs the constrained loss landscape. $\pi$ is given by a neural network whose outputs are mean $\mu_{\text{actor}}$ and covariance $\Sigma_{\text{actor}}$, where $\Sigma_{\text{actor}}$ primarily drives exploration; hence, we only use $\mu_{\text{actor}}$ for online control.
	\item \textbf{Soft constraints.}\quad When the state is very far from the goal, we have $$e^{-\frac{\norm{g - x_{t}^{i}}^2}{2 \sigma^2}} \approx 0 \quad \forall t=0,\ldots,N-1, i=1,\ldots,\hat{N}_s$$ meaning the $N$-step cost only accounts for constraint penalties $\epsilon$. This directly enables the agent to focus on short-term constraint satisfaction, and then consider long-term cost through the terminal value function. In contrast, weighing constraint violations can be cumbersome when using, for example, a quadratic stage cost. Alternatively, hard constraints may be used, which can result in the controller getting ``stuck'' trying to avoid violations, or becoming infeasible if improperly designed. This is illustrated in \cref{fig:cstr_rlmpc_profile} in \cref{subsec:cstr}. The proposed architecture avoids such issues. 
	\item \textbf{Scenario tree.}\quad $\hat{\psi}^i$ encompasses a set of scenarios, possibly different from those seen in the offline \ac{MDP}. For example, the policy \cref{eq:QMPC} might only factor in the extreme uncertainty realizations. As in \cref{eq:multiMPCobj}, we assume branching occurs only at the initial state $x^i_0 = s$, then predictions are performed over fixed scenarios for a short time horizon, leading to the terminal cost. 
\end{itemize}
To summarize, \cref{eq:QMPC} includes the core feature of scenario-based \ac{MPC}, which is planning over different realizations of a system model.
Moreover, it incorporates a high-level cost through the goal-conditioned \ac{RL} state value function, and a local Gaussian-shaped stage cost.
This stage cost enables the proposed value function-augmented \ac{MPC} agent to prioritize constraints at states far away from the goal.

\subsection{Interface with MPC theory}

We provide a sketch of how classical \ac{MPC} theory can be incorporated into the \ac{RL}-\ac{MPC} interface to ensure safety requirements.
So far, we have emphasized robustness in our formulation, specifically in a scenario-based sense due to its elegant relationship to dynamic programming, as discussed in \cref{subsec:multiMPC}.
Other safety aspects include stability, recursive feasibility, and optimality.

Stability typically relies on Lyapunov-based arguments using the \ac{MPC} cost as a Lyapunov function \cite{mayne2000Constrainedmodel}.
As such, it is necessary to exploit the structure of the stage cost and terminal value function in \ac{MPC}.
Desirable structures can be directly parameterized, such as a learnable convex stage cost \citep{seel2022ConvexNeurala}, or Lyapunov neural network as the terminal value function \cite{mukherjee2022NeuralLyapunov}.
The value function-augmented framework presented here is indeed compatible with these ideas due to its modularity; however, future work should study such problems in the context of the proposed architecture in \cref{eq:QMPC}.
This architecture has the unique property of isolating constraints in its objective for faraway states while avoiding feasibility issues.
If a more traditional setup is used instead, for example, with convex cost terms and hard constraints, then one may still apply Lyapunov arguments to arrive at recursive feasibility.

While optimality cannot be verified for general \acp{MDP} due to the intractability of dynamic programming, it is useful to view \ac{MPC} through the lens of suboptimal control \citep{bertsekas2005DynamicProgramming}.
Specifically, \ac{MPC} can be viewed as a rollout algorithm, yielding strong connections to policy iteration \citep{bertsekas2021rollout}.
In particular, the lookahead and optimization structures of \ac{MPC} enable it to guarantee improved control over a base policy.
Here, the base policy is generated through training \iac{RL} agent offline; the \ac{MPC} agent is able to improve upon it with the use of the \ac{RL}-based terminal value function in its cost.
This property is illustrated in \cref{fig:norm_profile}.


\section{Case studies}

We present three case studies.
The first two demonstrate the advantages of a goal-conditioned objective over a classical control objective in MPC in terms of handling nonlinear and high-dimensional system dynamics.
The first illustrates the goal-conditioned reward in \cref{eq:goalreward} in a nominal, nonlinear \ac{MPC} setup without \ac{RL}, compared against more traditional objectives.
The second demonstrates, theoretically and empirically, the advantages of a goal-conditioned objective through the lens of classical linear quadratic control both in terms of performance and scalability.
The third brings together all the elements discussed in this paper on \ac{RL} and \ac{MPC}: robustness, goal conditioning, and the combination of local-global values.
The corresponding code is available here: \url{https://github.com/NPLawrence/RL-MPC}.

\subsection{Example 1: Nominal goal-conditioned MPC}
\label{sec:dip}

This example focuses on a simplified version of the proposed policy structure in \cref{eq:QMPC}.
We consider nominal goal-conditioned \ac{MPC} applied to a double inverted pendulum.
That is, the \ac{MPC} policy is given a nominal physics model and does not consider system uncertainty, nor does it include \iac{RL} value function.
The purpose of this demonstration is to isolate the Gaussian-shaped reward in a planning context and to compare it against other \ac{MPC} agents.

The task is to apply force to a cart in order to bring the double inverted pendulum from its natural resting state to the upright position.
The goal can be formulated in terms of the angle of each link relative to the upright position.
Below is a summary of each \ac{MPC} agent.
\begin{itemize}
	\item \textbf{Expert.}\quad This is the formulation and implementation used in the benchmark example by \citet{fiedler2023DompcFAIR}, readily available in the authors' \texttt{\href{https://www.do-mpc.com}{do-mpc}} toolbox. The stage cost aims to maximize potential energy and minimize kinetic energy; it also includes a penalty term on changes to the actions to encourage ``smooth'' control.
	\item \textbf{Quadratic.}\quad The stage and terminal costs are $$\frac{1}{2}\left(\left( 1 - \cos(\theta_1)\right)^2 + \left( 1 - \cos(\theta_2)\right)^2\right),$$ where $\theta_1, \theta_2$ are the angles of the two links. The controller does not include a penalty term on the actions.
	\item \textbf{Goal-conditioned.}\quad The same setup as the quadratic formulation, but with the cost set to 
	\begin{equation}
		-e^{-\frac{1}{2}\left(\left( 1 - \cos(\theta_1)\right)^2 + \left( 1 - \cos(\theta_2)\right)^2\right)}.
	\label{eq:smoothreward}
	\end{equation}
\end{itemize}

\begin{figure}
	\includegraphics[width=3.3in]{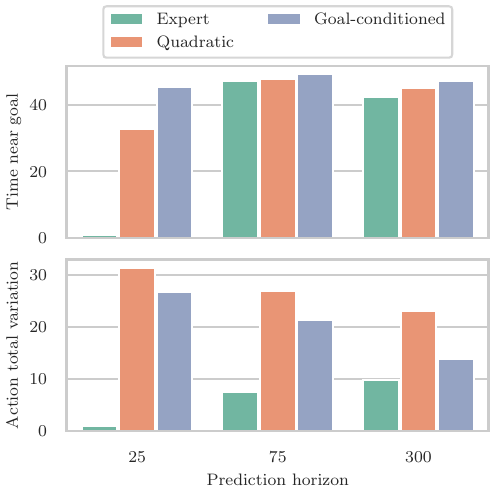}
	\caption{(Top) The goal-conditioned agent gives the most consistent performance in terms of time spent in the upright position. (Bottom) The expert agent is very efficient at solving the swing up task, whereas the quadratic agent is the most aggressive. The goal-conditioned agent becomes much more efficient with its actions as the prediction horizon increases.}
	\label{fig:dip_performance}
\end{figure}

We perform a sweep over three different prediction horizons.
\Cref{fig:dip_performance} summarizes the performance of each agent as follows: ``Time near goal'' is quantified using \cref{eq:smoothreward} with $\sigma^2 = 0.01$ (much more stringent than the goal-conditioned \ac{MPC} stage cost); ``Action total variation'' reports $\sum_{t=0}^{99} \norm{a_{t} - a_{t-1}}$ over the course of each $100$-time step experiment.

Based on \cref{fig:dip_performance}, the quadratic and goal-conditioned \ac{MPC} agents are able to solve the swing up task under the three prediction horizons, while the expert formulation fails when $N=25$.
For $N=75$, all three agents are approximately aligned in terms of time spent near the goal, but the expert agent does so with at most $1/3$ the action variation of the other two policies.
Planning very far into the future, the goal-conditioned agent creates slightly more separation from the other agents in time spent in the upright position.
However, its decrease in action variation is more noteworthy:
Across the three prediction horizons, the expert agent's action variation slowly increased, and the quadratic agent's slowly decreased.
In contrast, the goal-conditioned agent became roughly $45\%$ more efficient with its actions.
This illustrates the idea that a goal-conditioned objective does not react aggressively to large errors, like a quadratic objective.
Instead, a long-term view means the sensitivities of \cref{eq:smoothreward} ``light up'' most strongly to trajectories that bring the state to the goal.

\begin{figure}
	\includegraphics[width=3.3in]{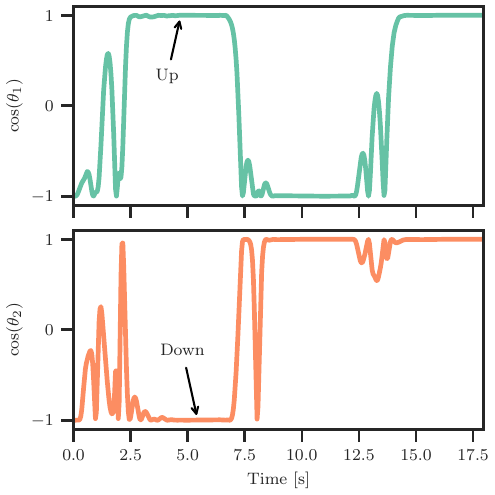}
	\caption{Starting from rest, the goal-conditioned \ac{MPC} agent is given a sequence of three different unstable equilibria to reach. A corresponding animation can be found here: \url{https://github.com/NPLawrence/RL-MPC}. This experiment was performed with $\sigma^2 = 0.5$ and $N=35$.}
	\label{fig:dip_profile}
\end{figure}

Our final experiment for this example showcases the goal-conditioned \ac{MPC} agent on the other two unstable equilibria of the double inverted pendulum.
While the expert agent is able to solve the swing up task, this is a secondary effect of its objective. 
In other words, it cannot readily be applied to the other equilibria.
\Cref{fig:dip_profile} shows a time profile of the angle trajectories as the goal-conditioned agent is directed to achieve different configurations.
This isolates and validates the use of a nonstandard goal-conditioned stage cost, independent of all the other machinery discussed in this paper.
The resulting \ac{MPC} agent is able to solve a complicated control problem efficiently.
However, we note the challenge of deploying \ac{MPC} alone with such an objective, namely, the potential inability to solve the task with a small variance value, which more accurately characterizes the goal as in \cref{eq:goalreward}.
This motivates the use of derivative-free optimization frameworks, such as \ac{RL}, for long-term goal-conditioned objectives, demonstrated in \cref{subsec:cstr}.

\subsection{Example 2: Classical control explanation of goal-conditioned policies}

This example serves two purposes.
First, we elucidate the proposed architecture in \cref{eq:QMPC} through a simplified setting.
In particular, we consider \ac{LTI} dynamics and show how \cref{eq:QMPC} relates to classical \ac{LQ} control in \cref{eq:nominalMPCobj}.
We show favorable performance gains of the goal-conditioned \ac{MPC} as the system dimension grows.
Second, we introduce \iac{RL}-trained value function into \cref{eq:QMPC} and study the online computational cost of the proposed architecture.
We demonstrate the real-time viability of incorporating \iac{RL}-learned value function into online \ac{MPC} agents.

\subsubsection{LQ-based lower bound on goal-conditioned objective}

Fix any stage cost $\ell$ and value function $V$, and let $\displaystyle\rho^{-1} = \frac{1-\gamma}{1 - \gamma^{N+1}}$.
We obtain the following:
\begin{equation}
\begin{aligned}
	e^{- \rho^{-1}\left( \sum_{t=0}^{N-1} \gamma^t \ell(x_t, u_t) + \gamma^N V(x_N) \right)} &<  \rho^{-1} \left( \sum_{t=0}^{N-1} \gamma^t e^{-\ell(x_t, u_t)} + \gamma^N e^{-V(x_N)} \right) \\
	&\approx (1 - \gamma) \lim_{N \to \infty} \sum_{t=0}^{N-1} \gamma^t \ind_{\{x_{t} = 0\}}.
\label{eq:jensen}
\end{aligned}
\end{equation}
The first line is an application of Jensen's inequality \cite{mordukhovich2013EasyPath}.
It follows due to convexity of the exponential with respect to the costs $\ell(x_t, u_t)$ and $V(x_N)$. The multiplier $\rho^{-1}$ ensures the summations are convex combinations; note we could also consider the finite-horizon average cost formulation instead of the discounted cost formulation.
The inequality is strict because of strict convexity over non-constant system trajectories.
The right-hand side represents an approximation of the goal-conditioned objective in the deterministic setting centered at the origin.

\begin{figure}
	\includegraphics[width=3.3in]{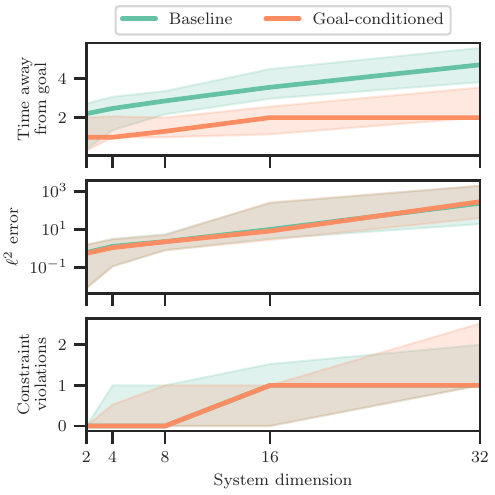}
	\caption{(Top) As the system dimension grows, the gap between the baseline controller and the goal-conditioned controller widens; (Middle/bottom) There is no appreciable difference in terms of other measures, such as $\ell^2$ error or constraint violations. Plots show median performance inside the $2.5-95$ percentile interval.}
\label{fig:scalability}
\end{figure}

The strict inequality in \cref{eq:jensen} is interesting because it relates some baseline objective on the left-hand side to an improved cost on the right-hand side without changing any of the fundamental ingredients, $\ell$, $V$, or the dynamic model.
Consider \ac{LTI} dynamics $x_{t+1} = A x_t + B u_t$ and quadratic stage cost $\ell(x,u) = x\transpose Q x + u\transpose R u$ with terminal cost $V(x) = x\transpose P x$ derived from \ac{LQR}.
We then see that maximizing the left-hand side of \cref{eq:jensen} is equivalent to the (discounted) \ac{LQR} objective.
Maximizing the right-hand side results in an improved goal-conditioned value over the baseline \ac{LQR} controller.
This is particularly beneficial when starting with a stabilizing baseline formulation, as it pushes the goal-conditioned counterpart on the right-hand side away from zero.

We validate the above discussion using \ac{MPC} alone; see \cref{app:LQ-GC} for full experimental details, as well as additional results studying the scalability of an \ac{RL}-learned value function inside \iac{MPC} agent.
We consider a random collection of linear systems across different dimensions.
For each one, we use a quadratic cost with $M = R = I$ and the resulting $P$ matrix from solving the \ac{LQR} problem. 
We use soft constraints, to ensure functional consistency in the $\ell$ terms used on both sides of \cref{eq:jensen}.
In the top plot of \cref{fig:scalability}, we accumulate the Gaussian-shaped reward in \cref{eq:goalreward} with a small variance ($\sigma^2 = 0.01$) and subtract it from the total number of time steps for each experiment.
It shows a growing gap between the baseline controller and the goal-conditioned controller as the system dimension increases; in other words, the goal-conditioned objective is less sensitive to dimensionality.
Because \ac{LQR} is not directly optimizing the goal-conditioned objective, we also report the $\ell^2$ error over each experiment.

An intuitive way to understand these results is as follows. 
For a given linear system at an arbitrary initial state, it is unlikely that either controller will monotonically decrease the error $x^T x$ (otherwise it would be a Lyapunov function). 
This means $\ell^2$ error will grow before it can decrease, which conflicts with the goal on the left-hand side of \cref{eq:jensen}.
Meanwhile, the goal-conditioned reward is insensitive to such behavior.
Indeed, when the initial state is far from the origin, it will not matter (in terms of cost) if the state briefly goes away from the origin.
Finally, the bottom plot tabulates constraint violations to show that the goal-conditioned controller does not suffer as a result of its more rapid approach to the goal.

\subsection{Example 3: Robust goal-conditioned policies for process control}
\label{subsec:cstr}

We study \iac{CSTR}, a common benchmark in process control, particularly in \ac{MPC} and learning-based applications \citep{fiedler2023DompcFAIR, bloor2024PCGymBenchmark, lee2001NeurodynamicProgramming, nejatbakhshesfahani2023Learningbasedstate}.
In our example, we use the model and parameters given by \citet{klatt1998Gainschedulingtrajectory}.
This is also the formulation used in the robust scenario-based \ac{MPC} benchmark by \citet{fiedler2023DompcFAIR}, readily available in the authors' \texttt{\href{https://www.do-mpc.com}{do-mpc}} toolbox.
For completeness, a short summary is given below, with the accompanying equations given in \cref{app:cstr}.


The \ac{CSTR} process is described by a fourth-order nonlinear ordinary differential equation.
The state variables are concentrations $c_A$ and $c_B$, reactor temperature $T_R$, and coolant temperature $T_K$.
The reaction $A \to B$ is controlled through the input variables $F$ (normalized inflow) and $\dot{Q}$ (heat removed by coolant).
Within this process are two additional reactions $B \to C$ and $A \to D$, forming byproducts $C$ and $D$.
Two key rate coefficients are considered uncertain. 
These rate terms depend exponentially on the reactor temperature $T_R$.
The uncertainty is characterized by two multipliers $\alpha$ and $\beta$: $\alpha$ characterizes uncertainty in the activation energy for the reaction $A \to D$, while $\beta$ characterizes uncertainty in the rate coefficient for the reaction $A \to B$.

\subsubsection{Robust offline training}

We train two agents:
\begin{itemize}
	\item \textbf{Nominal \ac{RL}.}\quad The environment does not contain any system uncertainty; only the true nominal parameter values are used.
	\item \textbf{Robust \ac{RL}.}\quad The environment is constructed with a branching process as described in \cref{eq:offlineMDP}. It assumes structural knowledge of the system dynamics and a range of possible values for $\alpha$ and $\beta$. This range is gridded and each value is considered equally likely, making \cref{eq:robustQ} the theoretical target.
\end{itemize}

The task for the agent is to control the concentration $c_B$ through the actions $F$ and $\dot{Q}$.
Therefore, for a desired concentration $c_{B}^{\text{goal}}$, the reward is defined as:
\begin{equation}
	r_g (s,a) = e^{-\frac{\left( c_{B}^{\text{goal}} - c_B \right)^2}{2 \sigma^2}}
\label{eq:cstrreward}
\end{equation}
with $\sigma^2 = 0.0001$.
The agents are trained using the soft actor-critic algorithm \citep{haarnoja2018Softactorcritic} and \ac{HER} for the replay buffer. See \cref{app:implementation} for further implementation details.

After training, we evaluate the agents based on how effectively they reach a novel goal.
This is measured as follows: $200$ initial states are randomly sampled; they are sampled from the constraint intervals used in \ac{MPC}, illustrated next in \cref{fig:cstr_rlmpc_profile} in \cref{subsec:cstr_rlmpc}. 
The agent is given $50$ time steps to reach the goal. 
The agent's effectiveness is quantified for the last $25$ time steps; this isolates steady state performance from the transient stage of each rollout.

\begin{figure}
	\includegraphics[width=3.3in]{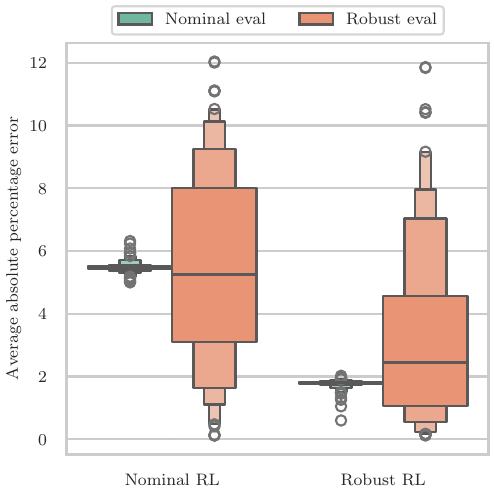}
	\caption{A ``boxen'' plot illustrating the distribution of absolute percentage error achieved by the nominal and robust \ac{RL} agents. The middle line is the median, the widest box contains $50\%$ of samples, then successively narrower boxes include additional $25\%$, $12.5\%, \ldots$ samples.}
	\label{fig:cstr_offlineagent}
\end{figure}

\Cref{fig:cstr_offlineagent} reports the average absolute percentage error over these final $25$ time steps.
The nominal evaluation means the aforementioned experiment is performed on the nominal plant without uncertainty; the robust evaluation samples and fixes parameter values from a set not used for the robust \ac{RL} training.
\Cref{fig:cstr_offlineagent} shows both the nominal and robust evaluation experiments leave about $5\%$ median offset over the final $25$ time steps, in contrast to the roughly $2\%$ of the robust \ac{RL} agent.
While it may be possible to fine-tune the hyperparameters to improve the nominal evaluation, it mainly serves as a reference.
The main takeaway is the robust evaluation, which shows a more modest dip by the robust \ac{RL} agent.
Overall, our primary motivation is to validate the robust \ac{RL} scheme, as we will incorporate it into the scenario-based \ac{MPC} framework next.

\subsubsection{Combining robust RL and scenario-based MPC}
\label{subsec:cstr_rlmpc}

\begin{figure}
	\includegraphics[width=3.3in]{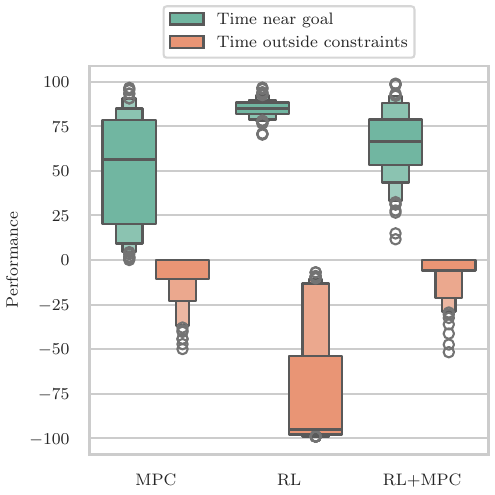}
	\caption{The robust \ac{MPC} agent shows excellent constraint satisfaction, but highly variable performance in terms of reaching the goal. The robust \ac{RL} agent has no knowledge of constraints, meaning it can quickly achieve its goals. The robust value function-augmented \ac{MPC} agent, dubbed ``RL+MPC'' balances the strengths of both.}
	\label{fig:cstr_rlmpc}
\end{figure}

We evaluate the performance of the value function-augmented scenario-based \ac{MPC} scheme in \cref{eq:QMPC}.
We compare it to a benchmark scenario-based \ac{MPC} scheme as well as the trained policy $\pi$ used to construct the terminal value $V$.
In \cref{fig:cstr_rlmpc}, we report the amount of time each agent spends near the goal as well as outside the constraints:
\begin{itemize}
	\item \textbf{Time near goal.}\quad We use the reward function in \cref{eq:cstrreward} with $\sigma^2=0.01$. We do not evaluate with the $\sigma^2$ value used for training because it is too sparse for all the comparisons.
	\item \textbf{Time outside constraints.}\quad We use the function $$e^{-\frac{\norm{s - \prox{\mathcal{X}}{s}}^2}{2 \sigma^2}} - 1,$$ also with $\sigma^2=0.01$, where $\text{prox}(\cdot)$ returns the closest point in the constraint set to the point of interest.
\end{itemize}
\Cref{fig:cstr_rlmpc} shows the sum of these quantities over $100$ time steps.
Like the previous section, we sample parameter configurations and initial states to collect these measurements for each agent.
\Cref{fig:full_cstr_rlmpc}, in \cref{app:cstr}, is a more comprehensive version of \cref{fig:cstr_rlmpc}, including nominal \ac{MPC} and other instances of the value function-augmented \ac{MPC} agent; we briefly note that the prediction horizon is the main tuning parameter in \cref{eq:QMPC}, which is a significant benefit of the proposed architecture.

\begin{figure}
	\includegraphics[width=3.3in]{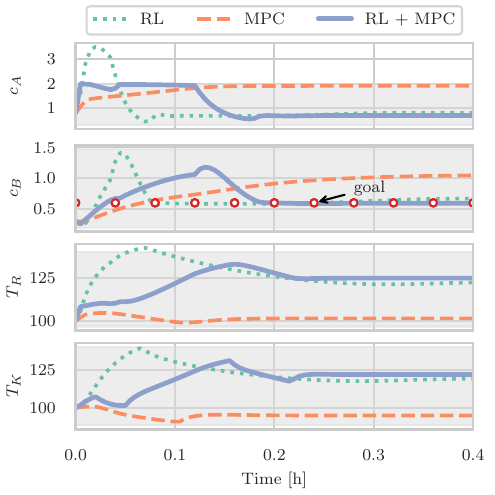}
	\caption{A time profile of each agent evaluated in \cref{fig:cstr_rlmpc}. The \ac{MPC} agent gets stuck satisfying the constraints but never achieving the goal, whereas the \ac{RL} agent immediately violates the constraints, but ultimately reaches the goal. The RL+MPC agent tames the \ac{RL} agent's trajectory, satisfying the constraints, while also eventually reaching the goal.}
	\label{fig:cstr_rlmpc_profile}
\end{figure}

\cref{fig:cstr_rlmpc} clearly indicates that the trained policy $\pi$ used to construct $V$ has no knowledge of the state constraints.
Indeed, it was trained in a global goal-driven fashion, meaning it comes as no surprise to see it almost always violates the constraints, indicated by a large mass at the bottom of \cref{fig:cstr_rlmpc}.
However, this comes with a significant payoff in terms of eventually achieving the goal, as the \ac{RL} agent spends more time overall near the goal than the \ac{MPC}-based agents. 
The robust \ac{MPC} agent shows good constraint satisfaction with its time outside constraints tapering away from zero.
However, there is wide variation in its time near the goal.
This manifests in trajectories where the agent does an excellent job at avoiding constraint violations, but is never able to reach the goal as a result.

Finally, the \ac{RL} $+$ \ac{MPC} agent, deploying \cref{eq:QMPC}, has similar or slightly tighter constraint satisfaction to the robust \ac{MPC} agent, but without the same extreme lows.
Overall, the time spent near the goal is more consistent (like the \ac{RL} agent), but with excellent constraint satisfaction.
A time profile of all three agents is shown in \cref{fig:cstr_rlmpc_profile}.
The initial state was chosen to illustrate the extreme behaviors of the \ac{MPC} and \ac{RL} agents that exist in \cref{fig:cstr_rlmpc}.
As indicated, the unconstrained \ac{RL} agent represents the quickest path to the goal, while the \ac{MPC} agent is only able to stick to the state constraints.
The \ac{RL} $+$ \ac{MPC} agent tempers the \ac{RL} agent's trajectory, hitting the upper bound of concentration $c_A$ just long enough until it can regulate $c_B$ effectively.

\section{Conclusions}
\label{sec:conclusion}

This paper advocated for a value function-centric perspective of \ac{RL} and \ac{MPC}, with the goal of enabling fluid discourse between both research areas.
To this end, we have demonstrated how value-based ideas enable a complementary control framework for solving \acp{MDP}.
Broadly speaking, \ac{RL} thrives at learning complex policies when it is able to freely explore its environment.
This is most readily achieved in simulation environments where concerns of safety are secondary; the benefit, however, is the ability to distill \emph{exploration} experience into high-level policies from simple rewards.
\Ac{MPC} represents another extreme, in which safety is at the forefront and achieved through repeated, online \emph{exploitation} of prior system knowledge, costs, and constraints.
We have shown that these differing perspectives \citep{banker2025LocalGlobalLearning} enable a single agent to utilize the strengths of both frameworks: \Iac{RL}-based terminal value function working in tandem with short-term \ac{MPC} planning.

While our combination of scenario-based planning and goal-conditioned learning contributes to this classical local-global view of \ac{RL} and \ac{MPC}, several challenges remain.
One of these is the possibility of mismatch between the distributions for the true environment and the system model.
While we did not assume an exact model is available, we did consider the environment to be well characterized by the system model uncertainty.
In principle, the robust \ac{RL} training setup could compensate for structural or parametric mismatch by including ``true'' data from the environment into its replay buffer alongside simulation data.
However, the proposed policy in \cref{eq:QMPC} would still contain a mismatched internal model.
Nonetheless, our proposed framework makes an initial step towards bringing together niche techniques from the vastly different \ac{RL} and \ac{MPC} communities.

\section*{Acknowledgement}
\label{sec:acknowledgements}
This material is based upon work supported by the U.S. Department of Energy, Office of Science, Office of Fusion Energy Sciences under award number DE‐SC0024472.
NPL would like to thank Professor Martha White for their constructive feedback.

\renewcommand*{\bibfont}{\footnotesize}
\bibliographystyle{elsarticle-num-names}
\bibliography{2024_RL_MPC}           

\appendix

\section{LQR value function}
\label{app:lqr}

The goal in \cref{eq:LQRobjective} is to \emph{regulate} a \emph{linear system} $x' = A x + B u$ as efficiently as possible according to a \emph{quadratic cost}. 
Although the \ac{LQR} problem contains an infinite number of decision variables, it turns out that the optimal solution is a static linear controller $u = -K x$.
This can be shown by combining the structure of the problem with Bellman's optimality equation.

\begin{enumerate}
	\item \textbf{Repurpose \cref{eq:bellmanQ}.}\quad Flipping signs, removing the expectation, plugging in the cost and dynamics equations, and finally minimizing both sides, we arrive at:
		\[\min_{u} Q^\star (x, u) = \min_{u} \left\{ x \transpose M x + u\transpose R u + \gamma \min_{u'} Q^\star (A x + B u, u') \right\} \]
	\item \textbf{Quadratic optimal cost.} ``Guess'' $\underset{u}{\min}\ Q^\star(x, u) = x\transpose P x$ for some symmetric $P$. (See below for an intuitive argument.) We then have 
		\[ x\transpose P x = \min_{u} \left\{ x \transpose M x + u\transpose R u + \gamma \left(A x + B u\right)\transpose P \left(A x + B u\right) \right\} \label{eq:bellmanLQR}\]
	\item \textbf{Solve for $u$.}\quad The right-hand side above can be solved by setting the gradient of the inside term equal to zero to find $u = -K x$, where
		\[K = \gamma \left(R + \gamma B\transpose P B\right)^{-1} B\transpose P A \]
	\item \textbf{Back-substitute.}\quad $K$ is expressed in terms of $P$. By plugging the solution $u = -K x$ back into \cref{eq:bellmanLQR} we arrive at the \ac{DARE}:
		\[P = M + \gamma A\transpose P A - \gamma^2 A\transpose P B \left(R + \gamma B\transpose P B \right)^{-1} B\transpose P A\]
\end{enumerate}

The \ac{DARE} is a tractable form of Bellman's optimality equation.
Like Bellman's equation, the desirability of the optimal solution in the \ac{DARE} depends on the discount factor.
For instance, as $\gamma \to 0$, the controller becomes degenerate, resulting in no control actions.
For an open-loop unstable system, this is clearly problematic.

Next, we illustrate why one would make the guess $\underset{u}{\min}\ Q^\star(x, u) = x\transpose P x$ in the first place (other than the simple fact that it works).
This is a two-step process:
\begin{enumerate}
	\item \textbf{Correspondence between linear controllers and quadratic value functions.}\quad Note that for any controller $\hat{K}$ with finite return, its value function is quadratic. Conversely, for any symmetric positive definite $\hat{P}$, solving
		\[
		\min_{u}\left\{ x\transpose M x + u\transpose R u + \left(A x + B u\right)\transpose \hat{P} \left(A x + B u\right) \right\}
		\]
		results in a linear controller.
	\item \textbf{The optimization problem in \cref{eq:LQRobjective} is lower bounded by the optimal linear controller.}\quad Define an surrogate objective over linear controllers:
\begin{equation}
\begin{aligned}
    &\underset{K}{\text{minimize}} && \sum_{t=0}^{\infty} \gamma^{t} \left( x_t\transpose M x_{t} + u_{t}\transpose R u_t \right) \\
    &\text{subject to } && x_{t+1} = A x_t + B u_t \\
    & && u_t = -K x_t
\end{aligned}
\label{eq:K-LQR}
\end{equation}
	Of course, \cref{eq:LQRobjective} lower bounds \cref{eq:K-LQR}.
	Showing the other direction starts with an auxiliary infinite-horizon objective:
\begin{equation}
\begin{aligned}
    &\underset{u_0, \ldots, u_{N-1}}{\text{minimize}} && \sum_{t=0}^{N-1} \gamma^{t} \left( x_t\transpose M x_{t} + u_{t}\transpose R u_t \right) + \text{cost}(\hat{K})\\
    &\text{subject to } && x_{t+1} = A x_t + B u_t
\end{aligned}
\label{eq:finitelqr}
\end{equation}
where $\text{cost}(\hat{K})$ is the cost of applying some linear controller $K$ after $N-1$ time steps.
Note that solving \cref{eq:finitelqr} results in a linear controller, meaning \cref{eq:K-LQR} lower bounds \cref{eq:finitelqr}.
Moreover, for $N = 1, 2, 3, \ldots$ the respective values in \cref{eq:finitelqr} decrease in $N$.
\end{enumerate}
Taken together, the \ac{LQR} objective is equivalent to optimizing over linear controllers, affirming the original choice of $\underset{u}{\min}\ Q^\star(x, u) = x\transpose P x$ in the Bellman equation.

\section{Additional results and details}

\subsection{Linear systems example}
\label{app:LQ-GC}

\textbf{Experimental setup.}\quad
We consider linear systems $x_{t+1} = A x_t + B u_t$ of dimensions $n = 2, 4, 8, 16, 32$.
Throughout, we take $B = I$ of suitable size.
To generate random $A$ matrices, we modify the formula by \citet{lawrence2024StabilizingReinforcement} for generating random stable matrices.
Sample a random matrix $M$ (here, we choose each entry from a uniform distribution), then update it as 
\[
M \leftarrow U \tanh(D) V\transpose,
\]
taking $U, D, V\transpose$ from the \ac{SVD} of the initial $M$ matrix.
Here, we clip and scale the $\tanh(D)$ term to allow for slightly unstable matrices (up to spectral radius $1.1$), and to better control the condition number of the system.
Next, update $M$ again via $$M \leftarrow L^{-1} M L,$$ where $L$ is a random lower triangular matrix with positive diagonal terms.

\begin{figure}
	\includegraphics[width=3.3in]{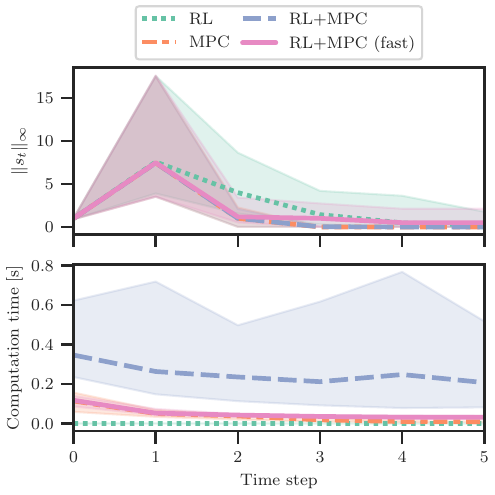}
	\caption{(Top) Direct goal-conditioned \ac{MPC} serves as a strong baseline across $10$ randomly chosen linear systems. While the \ac{RL} policy is reasonably performant, value-augmented \ac{MPC} serves as an improvement over the base \ac{RL} policy. Actions, not shown, are in $[-1, 1]$. (Bottom) We find the computation cost of value function-augmented \ac{MPC} (dashed) is not prohibitive for large network architectures and can be significantly improved through a local approximation (solid).}
	\label{fig:norm_profile}
\end{figure}

For each dimension, we randomly sample $10$ systems according to the above procedure ($50$ systems in total).
Each control strategy is evaluated on each system for $10$ randomly chosen initial states with entries in $[ -1, 1 ]$.
\cref{fig:scalability,fig:norm_profile} report median performance across all systems and initial states.

\textbf{\ac{RL} scalability.}\quad
We fix $n = 32$ and train an \ac{RL} agent for $500,000$ time steps for each of the $10$ random systems.
The agents are trained under the reward in \cref{eq:goalreward} with $\sigma^2 = 0.25$.
\cref{fig:norm_profile} shows state and computation profiles for four control strategies: the trained \ac{RL} policy, goal-conditioned \ac{MPC}, goal-conditioned \ac{MPC} with the trained \ac{RL} value function, and a ``fast'' variant.

The actor and critic architectures have over $90,000$ parameters each, so the terminal value function used in the \ac{MPC} agents has over $180,000$ parameters.
This is indeed overkill for the problem; one could certainly optimize the network architectures for more efficient online computation.
The main point is that the online optimization of the value function-augmented \ac{MPC} agent is not prohibitive.
While strategies to aid in the real-time deployment of advanced \ac{MPC} schemes lie beyond the scope of this paper, we offer an additional experiment.
We replace the value function with a Taylor approximation around the goal state.
This approach was demonstrated by \citet{salzmann2023RealtimeNeuralMPC} for incorporating neural network dynamic models into \ac{MPC} and is available in the L4CasADi package (see \cref{app:implementation}).
Under this approach, the online computation is comparable to standard goal-conditioned \ac{MPC} with minimal loss in performance compared to the full value function-augmented setup.
We note, however, this approach is based on a local approximation of the value function network, so we find the initial state cannot be too far from the goal, or the prediction horizon needs to be expanded.
Another possibility is to train \iac{MPC} structure that is known to be tractable for online control such that it approximates some more expensive but performant \ac{MPC} agent \citep{tamar2017learning, rickenbach2025ZipMPCCompressed}, such as the value function-augmented \ac{MPC}.

\subsection{CSTR example}
\label{app:cstr}

\textbf{CSTR system description.}\quad
Refer to \cref{tab:cstr} for parameter values in the following system model:
\begin{align}
	\dot{c}_{A} &= F \left( c_{A,0} - c_A \right) - k_1 c_A - k_3 c^2_A \\
	\dot{c}_{B} &= -F c_B + k_1 c_A - k_2 c_B \\
	\dot{T}_{R} &= \frac{k_1 c_A H_{R, ab} + k_2 c_B H_{R, bc} + k_3 c^2_A H_{R, ad}}{- \kappa C_p} + F \left( T_{\text{in}} - T_R \right) \frac{K_w A_R \left( T_K - T_R \right)}{\kappa C_p V_R} \\
	\dot{T}_{K} &= \frac{\dot{Q} + K_w A_R T_{\text{dif}}}{m_k C_{p,k}},
\end{align}
where
\begin{align}
	k_1 &= \beta k_{0,ab}\ \text{exp} \left( \frac{-E_{A, ab}}{T_R + 273.15} \right) \\
	k_2 &= k_{0, bc}\ \text{exp} \left( \frac{-E_{A,bc}}{T_R + 273.15} \right)\\
	k_3 &= k_{0, ad}\ \text{exp} \left( \frac{-\alpha E_{A,ad}}{T_R + 273.15} \right).
\end{align}
$\alpha, \beta$ are uncertainty parameters, with nominal values of $1.0$. For robust \ac{MPC}, the extreme values for $\alpha$ are $0.95$ and $1.05$; for $\beta$ they are $0.9$ and $1.1$.
For robust \ac{RL} training, we take these intervals and grid them into $10$ evenly spaced values.

\begin{table*}[tbp]
\caption{Certain parameters in the CSTR model.}
\begin{center}
\begin{tabular}{ll|ll}
\toprule
$k_{0,ab}$ & $1.287 \cdot 10^{12}\ \text{h}^{-1}$ & $\kappa$ & $0.9342$ kg/l \\
$k_{0, bc}$ & $1.287 \cdot 10^{12}\ \text{h}^{-1}$ & $C_p$ & $3.01$ kJ/kg K \\
$k_{0, ad}$ & $9.043 \cdot 10^9$ l/mol h & $C_{p,k}$ & $2.0$ kJ/kg K \\
$R$ & $8.3144621 \cdot 10^{-3}$ & $A_R$ & $0.215$ $\text{m}^2$ \\
$E_{A, ab}$ & $9758.3 \cdot R$ kJ/mol & $V_R$ & $10.01$ l\\
$E_{A, bc}$ & $9758.3 \cdot R$ kJ/mol & $m_k$ & $5.0$ kg \\
$E_{A, ad}$ & $8560.0 \cdot R$ kJ/mol & $T_{\text{in}}$ & $130.0$ $^{\circ}$C \\
$H_{R, ab}$ & $4.2$ kJ/molA & $K_w$ & $4032.0$ kJ/h $\text{m}^2$ K \\
$H_{R, bc}$ & $-11.0$ kJ/molB & $c_{A,0}$ & $5.1$ mol/l \\
$H_{R, ad}$ & $-41.85$ kJ/molA & & \\
\bottomrule
\end{tabular}
\end{center}
\label{tab:cstr}  
\end{table*}

\textbf{Tuning the value function-augmented \ac{MPC} agent.}\quad
The main tuning parameters in \cref{eq:QMPC} are the prediction horizon and the variance in the Gaussian-shaped reward.
One could, in principle, also manipulate the uncertainty set and the state-action constraints. 
We consider these to be fixed.
Moreover, one could explore the possibility of fine-tuning the \ac{RL} agent online:
If the uncertainty set is not sufficiently accurate, then the \ac{RL} agent could adapt its policy to online data; however, this complicates the interplay between the global \ac{RL} value function and the now-inaccurate local replanning.
This was also discussed in \cref{sec:conclusion} and is a topic for future research.

\begin{figure}
	\includegraphics[width=3.3in]{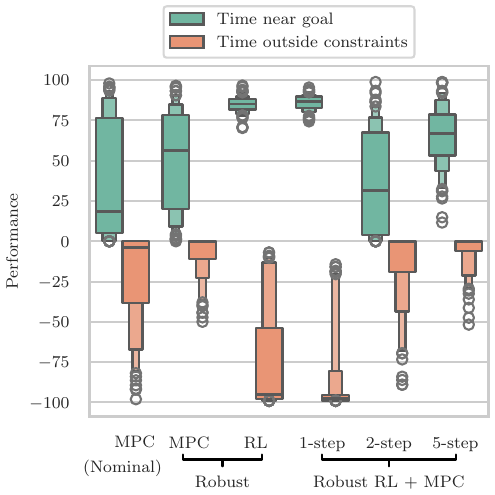}
	\caption{A comprehensive version of \cref{fig:cstr_rlmpc}.}
	\label{fig:full_cstr_rlmpc}
\end{figure}

We fixed the variance parameter to be $\sigma^2 = 0.25^2$.
This was fixed somewhat arbitrarily.
Mainly, it is larger than the value used by the \ac{RL} agent ($\sigma^2 = 0.0001$), but still small relative to the range of the target variable ($0.1-2.0$).
The more important parameter of the value function-augmented \ac{MPC} agent is the prediction horizon, as the terminal value function is concerned with performance, while the planning steps are concerned with constraint satisfaction.
\Cref{fig:full_cstr_rlmpc} shows the effect of varying the prediction horizon.
Overall, increasing the prediction horizon leads to better constraint satisfaction.
We only needed to set $N=5$ to get roughly the same result as the benchmark \ac{MPC} with $N=20$.
Finally, only having the prediction horizon as the main tuning knob is a significant advantage, especially when one can start from $N=0$ as the best option for pure performance.

\section{Implementation details}
\label{app:implementation}

All experiments were performed on an Apple M3 Pro with 18 GB of RAM.
The software implementation of the ideas presented in this paper relied heavily on \texttt{\href{https://www.do-mpc.com}{do-mpc}} \citep{fiedler2023DompcFAIR}, \texttt{\href{https://docs.cleanrl.dev}{CleanRL}} \citep{huang2022CleanRLHighquality}, and \texttt{\href{https://github.com/Tim-Salzmann/l4casadi}{L4CasADi}} \citep{salzmann2024LearningCasADi}.
We outline how these tools were used in this work.
The corresponding code is available here: \url{https://github.com/NPLawrence/RL-MPC}.

\begin{itemize}
	\item \textbf{do-mpc} is an all-in-one toolbox for nonlinear and scenario-based \ac{MPC}, including simulation, state estimation, and data management tools. It has a modular structure, which enabled us to integrate it into \iac{RL} pipeline. More specifically, we created a Gym environment for \ac{RL} training by using the do-mpc simulator to generate the state transitions under uncertain system parameters. At deployment time, we created the value function-augmented \ac{RL} agent in \cref{eq:QMPC} by specifying the value function as the terminal cost, specifying soft constraints, and defining a custom stage cost, all subject to a set of uncertainty parameters.
	\item \textbf{CleanRL} is a library of single-file implementations of deep \ac{RL} algorithms. This structure streamlines the process of building a custom pipeline involving the do-mpc-based Gym environment and the option to evaluate a nonstandard policy. In our experiments, we did not stray too far from the default hyperparameter settings. While we do not claim this is general advice, we found it useful to increase the batch size and use smooth activation functions. This is because the reward is often sparse in our setting, so a larger batch size improves the quality of the underlying expectation estimates. Smooth activation functions, such as ``swish'' \citep{ramachandran2017SwishSelfGated}, are also helpful for online \ac{MPC} optimization.
	\item \textbf{L4CasADi} is a package that enables the integration of PyTorch models with CasADi. Since do-mpc is an API for CasADi \citep{Andersson2019}, L4CasADi was essential for using the \ac{RL}-learned value function as a terminal cost in \cref{eq:QMPC}. do-mpc offers some functionality for integrating ML models, but L4CasADi is a much more flexible option.
\end{itemize}

\end{document}

%% file: acronyms.tex
\DeclareAcronym{AI}{short = AI,
	long = artificial intelligence}
\DeclareAcronym{ML}{short = ML,
	long = machine learning,
	short-indefinite = an}

\DeclareAcronym{SVD}{short = SVD, long = singular value decomposition}
\DeclareAcronym{CCA}{short = CCA,
	long = canonical correlation analysis}	
\DeclareAcronym{FA}{short = FA,
	long = factorial analysis}
\DeclareAcronym{GMM}{short = GMM,
	long = Gaussian mixture model}	
\DeclareAcronym{ICA}{short = ICA,
	long = independent component analysis}	
\DeclareAcronym{LARS}{short = LARS,
	long = least-angle regression}	
\DeclareAcronym{LASSO}{short = LASSO,
	long = least absolute shrinkage and selection operator}	
\DeclareAcronym{LR}{short = LR,
	long = logistic regression}
\DeclareAcronym{PCA}{short = PCA,
	long = principal component analysis}
\DeclareAcronym{PLS}{short = PLS,
	long = partial least squares}	
\DeclareAcronym{RBC}{short = RBC,
	long = reconstruction-based contribution}

\DeclareAcronym{ANFIS}{short = ANFIS,
	long = adaptive network fuzzy inference system}
\DeclareAcronym{ANN}{short = ANN,
	long = artificial neural network,
	short-indefinite = an}	
\DeclareAcronym{BN}{short = BN,
	long = Bayesian network}	
\DeclareAcronym{CNN}{short = CNN,
	long = convolutional neural network}	
\DeclareAcronym{DNNE}{short = DNNE,
	long = decorrelated neural network ensemble}		
\DeclareAcronym{DNN}{short = DNN,
	long = deep neural network}	
\DeclareAcronym{ELM}{short = ELM,
	long = extreme learning machine}
\DeclareAcronym{GAN}{short = GAN,
	long = generative adversarial network}
\DeclareAcronym{GPR}{short = GPR,
	long = Gaussian process regression}	
\DeclareAcronym{GRNN}{short = GRNN,
	long = general regression neural network}	
\DeclareAcronym{MLP}{short = MLP,
	long = multilayer perceptron,
	short-indefinite = an}
\DeclareAcronym{RBFNN}{short = RBFNN,
	long = radial basis function neural network,
	short-indefinite = an}
\DeclareAcronym{RNN}{short = RNN,
	long = recurrent neural network,
	short-indefinite = an}
\DeclareAcronym{RT}{short = RT,
	long = regression tree,
	short-indefinite = an}			
\DeclareAcronym{RVM}{short = RVM,
	long = relevance vector machine,
	short-indefinite = an}		
\DeclareAcronym{SFA}{short = SFA,
	long = slow feature analysis}	
\DeclareAcronym{SVM}{short = SVM,
	long = support vector machine}
\DeclareAcronym{TL}{short = TL,
	long = transfer learning}	
\DeclareAcronym{VAE}{short = VAE,
	long = variational autoencoder}	
\DeclareAcronym{WNN}{short = WNN,
	long = wavelet neural network}				

\DeclareAcronym{RL}{short = RL, long = reinforcement learning, short-indefinite = an}
\DeclareAcronym{A3C}{short = A3C,
	long = asynchronous advantage actor-critic}	
\DeclareAcronym{ADP}{short = ADP,
	long = approximate dynamic programming}
\DeclareAcronym{IQC}{short = IQC, long = integral quadratic constraint}
\DeclareAcronym{BIBO}{short = BIBO, long = {bounded-input, bounded-output}}
\DeclareAcronym{SISO}{short = SISO, long = {single-input, single-output}}
\DeclareAcronym{MIMO}{short = MIMO, long = {multiple-input, multiple-output}}
\DeclareAcronym{DDPG}{short = DDPG,
	long = deep deterministic policy gradient}
\DeclareAcronym{DQN}{short = DQN,
	long = deep $Q$-network}
\DeclareAcronym{HJB}{short = HJB,
	long = {Hamilton-Jacobi-Bellman}}
\DeclareAcronym{MPC}{short = MPC, 
	long = model predictive control,
	long-plural-form = model predictive controllers,
	short-indefinite = an}		
\DeclareAcronym{PI2}{short = $\text{PI}^2$,
	long =policy improvement with path integrals}
\DeclareAcronym{PID}{short = PID, 
	long = proportional-integral-derivative}
\DeclareAcronym{PI}{short = PI, 
	long = proportional-integral}
\DeclareAcronym{PPO}{short = PPO,
	long = proximal policy optimization}
\DeclareAcronym{REINFORCE}{short = REINFORCE,
	long = {\emph{RE}ward Increment $=$ \emph{N}onnegative \emph{F}actor $\times$ \emph{O}ffset \emph{R}einforcement $\times$ \emph{C}haracteristic \emph{E}ligibility}}
\DeclareAcronym{RTO}{short = RTO,
	long = real-time optimization,
	short-indefinite = an}	
\DeclareAcronym{SAC}{short = SAC,
	long = soft actor-critic}	
\DeclareAcronym{TD3}{short = TD3,
	long = twin-delayed DDPG}	
\DeclareAcronym{HER}{short = HER,
	long = hindsight experience replay}

\DeclareAcronym{GP}{short = GP,
	long = Gaussian process,
	long-plural-form = Gaussian processes}
\DeclareAcronym{RBF}{short = RBF,
	long = radial basis function,
	short-indefinite = an}	
\DeclareAcronym{SAE}{short = SAE,
	long = sparse autoencoder}	
\DeclareAcronym{DBN}{short = DBN,
	long = deep belief network}	
\DeclareAcronym{LSTM}{short = LSTM,
	long = long short-term memory}	
\DeclareAcronym{KL}{short = KL,
	long = Kullback-Leibler}
\DeclareAcronym{MDP}{short = MDP,
	long = Markov decision process,
	long-plural-form = Markov decision processes,
	short-indefinite = an}
\DeclareAcronym{LQR}{short = LQR, 
	long = linear quadratic regulator}
\DeclareAcronym{LQ}{short = LQ, 
	long = linear quadratic}
\DeclareAcronym{DARE}{short = DARE, 
	long = discrete algebraic Riccati equation}
\DeclareAcronym{LTI}{short = LTI, 
	long = linear time-invariant,
	short-indefinite = an}
\DeclareAcronym{GPS}{short = GPS,
	long = guided policy search}	
\DeclareAcronym{GRU}{short = GRU,
	long = gated recurrent unit}	
\DeclareAcronym{ESN}{short = ESN,
	long = echo state network}	
\DeclareAcronym{ENN}{short = ENN,
	long = Elman neural network}	
	
\DeclareAcronym{CSTR}{short = CSTR,
	long = continuous stirred tank reactor}